%% file: main.tex
\documentclass[a4paper, 11pt, dvipsnames, table]{article}
\usepackage{jcap}
\usepackage{natbib}
\setcitestyle{aysep={}}
\usepackage{xcolor}

\renewenvironment{thebibliography}[1]{%
\begin{oldthebibliography}{#1}%
\small%
\linespread{0.9}\selectfont
\raggedright%
\setlength{\itemsep}{10pt plus 0.2ex minus 0.05ex}%
}%
{%
\end{oldthebibliography}%
}

\usepackage{preamble}

\begin{document}

\title{Map-level baryonification: unified treatment of weak lensing two-point and higher-order statistics}

\author[1, 3]{Alan Junzhe Zhou}
\author[2, 3]{Marco Gatti}
\author[2, 3]{Dhayaa Anbajagane}
\author[2,3,4]{Scott Dodelson}
\author[5,6]{Matthieu Schaller}
\author[5]{Joop Schaye}
\affiliation[1]{Department of Physics, University of Chicago, Chicago, IL 60637, USA}
\affiliation[2]{Department of Astronomy and Astrophysics, University of Chicago, Chicago, IL 60637, USA}
\affiliation[3]{Kavli Institute for Cosmological Physics, University of Chicago, Chicago, IL 60637, USA}
\affiliation[4]{Fermi National Accelerator Laboratory, P.O. Box 500, Batavia, IL 60510, USA}
\affiliation[5]{Leiden Observatory, Leiden University, PO Box 9513, 2300 RA Leiden, the Netherlands}
\affiliation[6]{Lorentz Institute for Theoretical Physics, Leiden University, PO box 9506, NL-2300 RA Leiden, the Netherlands}

\emailAdd{ajzhou@uchicago.edu}
\abstract{
Precision cosmology benefits from extracting maximal information from cosmic structures, motivating the use of higher-order statistics (HOS) at small spatial scales. However, predicting how baryonic processes modify matter statistics at these scales has been challenging. The baryonic correction model (BCM) addresses this by modifying dark-matter-only simulations to mimic baryonic effects, providing a flexible, simulation-based framework for predicting both two-point and HOS. We show that a 3-parameter version of the BCM can jointly fit weak lensing maps' two-point statistics, wavelet phase harmonics coefficients, scattering coefficients, and the third and fourth moments to within 2\% accuracy across all scales $\ell < 2000$ and tomographic bins for a DES-Y3-like redshift distribution ($z \lesssim 2$), using the \flamingo{} simulations. These results demonstrate the viability of BCM-assisted, simulation-based weak lensing inference of two-point and HOS, paving the way for robust cosmological constraints that fully exploit non-Gaussian information on small spatial scales.
}

\maketitle
\flushbottom

\input{1_intro}
\input{2_BCM}
\input{3_sim}
\input{4_stats}
\input{5_method}
\input{6_results}
\input{7_conc}

\section*{Acknowledgments}
AZ is supported by Jane Street through the Jane Street Graduate Research Fellowship. MG is supported by funds provided by the Kavli Institute for Cosmological Physics at the University of Chicago through an endowment from the Kavli Foundation. DA is supported by the National Science Foundation Graduate Research Fellowship under Grant No. DGE 1746045. This work was supported by FermiForward Discovery Group, LLC under Contract No. 89243024CSC000002 with the U.S. Department of Energy, Office of Science, Office of High Energy Physics. This work used the DiRAC@Durham facility managed by the Institute for
Computational Cosmology on behalf of the STFC DiRAC HPC Facility
(\url{www.dirac.ac.uk}). The equipment was funded by BEIS capital funding via
STFC capital grants ST/K00042X/1, ST/P002293/1, ST/R002371/1 and ST/S002502/1,
Durham University and STFC operations grant ST/R000832/1. DiRAC is part of the
National e-Infrastructure.

\section*{Data Availability}

The \textsc{BaryonForge} pipeline used in this work is publicly available at \url{https://github.com/DhayaaAnbajagane/BaryonForge}. The \flamingo{} simulations are available upon request by contacting Joop Schaye at \texttt{schaye@strw.leidenuniv.nl}. The total matter power spectra from the \flamingo{} simulations have been publicly released and are available at \url{https://flamingo.strw.leidenuniv.nl}.

\clearpage
\newgeometry{left=1.2in,right=1.2in,top=1in,bottom=1in}
\bibliographystyle{mnras}
\bibliography{References,References_alan}
\restoregeometry

\input{main_appendix}

\label{lastpage}
\end{document}

%% file: 1_intro.tex
\section{Introduction}
Weak gravitational lensing directly probes the total matter distribution in the Universe, making it a powerful cosmological tool. Unlike galaxy clustering, which is biased by galaxy-matter relationships, weak lensing is sensitive to all matter, both dark and baryonic. Measurements of two-point weak lensing statistics have provided some of the tightest constraints on cosmological parameters, particularly $S_8 = \sigma_8 \sqrt{\Omega_m/0.3}$, which captures the degeneracy between the amplitude of matter fluctuations ($\sigma_8$) and matter density ($\Omega_m$) \citep{Amon2022,Secco2022,Li2023_,Dalal2023,des_kids_reanalysis, Anbajagane:2025:DECADECosmo, Wright2025}.

There is growing interest in extracting non-Gaussian features from the lensing map, both at the field level \citep{alsingCosmologicalParametersShear2017,loureiroAlmanacWeakLensing2022, boruahMapbasedCosmologyInference2023a, zhouAccurateFieldlevelWeak2024, daiMultiscaleFlowRobust2024,zhouHamiltonianPostBornThreedimensional2024} and via higher-order statistics (HOS). In particular, many HOS have been successfully applied to data have shown significant gains in constraining power (e.g., \citealt{Fluri2019,Gatti2022MomentsDESY3,jeffreyDarkEnergySurvey2024a,Marques2024,Novaes2024,gattiDarkEnergySurvey2024,Cheng2025}). They are also expected to improve our constraining power on extended cosmological models, such as those with primordial non-Gaussianities \citep{Anbajagane2023Inflation} or modified theories of gravity \citep{Liu:2016:MG, Davies:2024:MG}. Unlike two-point statistics, however, most HOS lack analytical predictions and rely on realistic weak lensing simulations for modeling.

Weak lensing observables -- both standard two-point and HOS -- probe \comments{low redshifts ($z < 1$) and small physical scales ($r < 10 \,\Mpc$),} making them particularly sensitive to non-linear structure formation and astrophysical processes such as supernova and active galactic nucleus feedback, collectively referred to as baryonic feedback. These astrophysical processes alter both the distribution and thermodynamics of baryons, and affect the dark matter distribution through gravitational coupling  (e.g., \citealt{Gnedin2004AdiabaticContraction,Duffy2010BaryonDmProfileDensity,vandaalenEffectsGalaxyFormation2011,semboloniEffectBaryonicFeedback2013,Anbajagane2022Baryons,Shao2022Baryons,mccarthyFLAMINGOProjectRevisiting2023,broxtermanFLAMINGOProjectBaryonic2024,schallerFlamingoProjectBaryon2025,schallerAnalyticRedshiftindependentFormulation2025}). These effects are important on small and intermediate scales, extending into the quasi-linear regime ($\lesssim 5\, \Mpc$), where they contribute significantly to the lensing signal. If unaccounted for, baryonic effects can bias cosmological inferences \citep{semboloniQuantifyingEffectBaryon2011,Yoon2019,Gatti2022MomentsDESY3,amonNonlinearSolution82022,Anbajagane:2023:CDFs}. For example, the observed discrepancy in $ S_8 $ between weak lensing and CMB measurements could arise from such mis-modeling rather than a fundamental deviation from the standard cosmological model \comments{\citep[e.g.,][]{amonNonlinearSolution82022}}. While a very conservative approach is to remove scales affected by baryons or to implement projection schemes that mitigate their impact \citep{Amon2022, Secco2022, Piccirilli2025, Truttero2025}, accurate modeling remains the preferred solution.

On the modeling side, the baryonic imprints on the matter density field are typically studied using high-resolution cosmological hydrodynamical simulations \citep[see][for a review]{Vogelsberger2020Hydro}. However, there is significant freedom in defining the ``subgrid’’ physics in these simulations -- processes like star formation and collisional cooling that occur below the simulation’s resolution scale. Different but \textit{equally well-motivated} subgrid prescriptions result in a wide range of predictions for the distribution of galaxy properties \citep[e.g.,][]{velliscigImpactGalaxyFormation2014,vandaalenContributionsMatterOutside2015,debackereImpactObservedBaryon2020,Anbajagane2020StellarProp,Lim2021GasProp,Lee2022rSZ,Stiskalek2022TNGHorizon,Anbajagane2022Baryons,Anbajagane2022GalaxyVelBias,Shao2022Baryons,Gebhardt2023CamelsAGNSN}, and consequently for the baryonic imprints on the non-linear matter power spectrum \citep[e.g.,][]{vandaalenEffectsGalaxyFormation2011,Chisari2018BaryonsPk,vandaalenExploringEffectsGalaxy2020,amonNonlinearSolution82022,schallerAnalyticRedshiftindependentFormulation2025}. While it is computationally unfeasible to explore broad cosmological and baryonic parameter spaces with these simulations,\footnote{\comments{Simulation suites like CAMELS \citep{Paco:2021:Camels} provide a wide variety of simulated predictions spanning this broad parameter space. However, these simulations are performed in cosmologically small volumes ($V \sim (25 \,\Mpc/h)^3$) and so cannot be reliably used to explore the impact of baryons on statistics summarizing the large-scale matter distribution.}} they are nonetheless essential for validating and calibrating simpler, more practical parametric models of baryonic effects.

Several parametric models have been developed to flexibly describe the impact of baryons on weak lensing observables.
A well-known example is HMCode2020 \citep{Mead2020:HMx}, an augmented halo model designed to accurately predict the non-linear matter power spectrum across a wide range of cosmologies. To account for baryonic feedback, HMCode2020 modifies the concentration–mass relation through a parametric model that accounts for mass and redshift-dependence and is fitted to the matter power spectra of the BAHAMAS simulations \citep{mccarthyBahamasProjectCalibrated2017}. 
More recent examples include the power spectrum emulator trained on the FLAMINGO simulations \citep{schallerFlamingoProjectBaryon2025} and a wavenumber-dependent rescaling of the difference between the non-linear and linear gravity-only power spectra \citep{amonNonlinearSolution82022,schallerAnalyticRedshiftindependentFormulation2025}.
Another successful approach is the baryonic correction model (BCM), a phenomenological method that mimics baryonic effects by perturbing the density field of dark matter-only (DMO) simulations via post-processing \citep{schneiderNewMethodQuantify2015,schneiderQuantifyingBaryonEffects2019,anbajaganeMaplevelBaryonificationEfficient2024}. These modified outputs can then be used to build emulators of the matter power spectrum that include such baryonic imprints. The BCM has demonstrated sufficient flexibility to reproduce a wide range of feedback scenarios seen in hydrodynamical simulations \citep{schneiderQuantifyingBaryonEffects2019,aricoModellingLargescaleMass2020,giriEmulationBaryonicEffects2021}, and has already been applied to analyze two-point statistics from wide-field surveys \citep{Chen2023,Arico2023,bigwoodWeakLensingCombined2024}.

Most of the recipes described above have been extensively tested in the context of two-point statistics. In principle, the BCM approach -- via its use of N-body simulations -- allows the construction of emulators for any observable, including HOS. However, the robustness of the model predictions for beyond two-point functions, in the context of analyzing current and future weak lensing datasets, remains largely unproven. Some studies have begun to explore this direction: \cite{semboloniEffectBaryonicFeedback2013} proposed a halo-based model which corrects the baryonic effects on the two and three-point functions. \cite{Arico2022bispectrum} showed that the BCM can simultaneously fit the matter power spectrum and bispectrum of hydrodynamical simulations for snapshots at different redshifts. \cite{lee2023} compared weak lensing peak counts from the full hydrodynamical simulation IllustrisTNG to those from a baryon-corrected version of its dark matter-only (DMO) counterpart, IllustrisTNG-Dark. They found that the BCM provided an insufficient fit when the noise levels match those expected from future surveys like LSST and Euclid. However, their analysis relied on BCM parameters calibrated solely to the power spectrum, whereas a joint fit that included peak counts could have improved performance. Moreover, they did not consider other types of HOS. More recently, \citet{anbajaganeMaplevelBaryonificationEfficient2024} performed a joint fit of second-, third-, and fourth-moments of the matter density and gas pressure fields measured in IllustrisTNG, finding statistical agreement with the model on scales above $\sim 1\, \Mpc$ and across multiple redshifts. However, this test was also limited to projected slices at fixed redshifts and did not validate against realistic weak lensing observables.

We show that the BCM is sufficiently flexible for simulation-based analyses by applying it to the \flamingo{} simulations, a state-of-the-art suite that features multiple baryonic feedback models, matching DMO runs, and is calibrated against observed $z=0$ galaxy stellar mass function and the galaxy cluster gas fractions \citep{Schaye2023, Kugel2023}. We baryonify the DMO simulations, generate realistic weak lensing maps, and build an emulator for both two-point and higher-order statistics (HOS) — including scattering transform coefficients \citep{chengNewApproachObservational2020}, wavelet phase harmonics \citep{allysNewInterpretableStatistics2020}, and third- and fourth-order moments. The emulator accurately recovers the summary statistics of the hydrodynamical runs, and we show that appropriate BCM parameter choices can jointly fit all statistics across the \flamingo{} feedback models down to $\ell \approx 2000$.

The paper is organized as follows. \Sec\ref{sec:bary} discusses the BCM model. \Sec\ref{sec:sim} describes the \flamingo{} simulations and the DES Y3-like tomographic weak lensing convergence maps. \Sec\ref{sec:stat} defines the relevant HOS and how they are measured in simulations. \Sec\ref{sec:method} describes the procedure for fitting the BCM model to the \flamingo{} feedback variants, and \Sec\ref{sec:result} presents the results. We conclude in \Sec\ref{sec:conc}. Supporting figures and tables are provided in the Appendix.

%% file: 2_BCM.tex
\section{Baryonification of simulations with \textsc{BaryonForge}}\label{sec:bary}
The BCM introduced in \citet{schneiderNewMethodQuantify2015} is a phenomenological framework that does not require specifying any small-scale subgrid physics. The method is parameterized by pairs of halo density profiles: one profile representing a halo in a DMO simulation, and another representing the halo in a hydrodynamical simulation (DMB). The latter includes contributions from dark matter (DM) and baryons (B), in the form of stars, gas, satellite galaxies, and dark matter. There exist different models for the exact form and parameterization of these profile components \citep{schneiderNewMethodQuantify2015, schneiderQuantifyingBaryonEffects2019, aricoModellingLargescaleMass2020, debackereImpactObservedBaryon2020, Pandey2024godmax}. Here we use the model of \citet{giriEmulationBaryonicEffects2021}, which is a slightly modified version of the model in \citet{schneiderQuantifyingBaryonEffects2019}. We refer the reader to those works for detailed discussions. Briefly, the total matter in the DMO case follows a Navarro–Frenk–White (NFW) profile \citep{navarroStructureColdDark1996,navarroUniversalDensityProfile1997} with a power-law cut-off at large scales, $r > 4 \Rtwohc$ \citep{Oguri:2011:truncation}. The stellar density follows an exponential profile, while the gas density follows a modified generalized NFW (gNFW) profile. The collisionless matter (dark matter and satellite galaxies) is modeled as a modified NFW profile that accounts for adiabatic relaxation due to the presence of the star and gas components. In our implementation, each halo is assigned a NFW concentration using the semi-analytic model of \citet{Diemer2015Concentration}.

The DMO and DMB profiles can be expressed via the quantity $M(r)$, the mass enclosed within a radius $r$, which is a monotonically increasing function. The DMO and DMB enclosed mass profiles are then combined to compute a displacement function,
\begin{equation}
    \Delta d(r) = M_{\rm DMB}^{-1}\left(M_{\rm DMO}(r)\right) - r.
\end{equation}
where the functions above are implicitly also functions of the halo mass, $M_{\rm 200c}$. Here, the function $M_{\rm DMB}^{-1}$ takes a mass as its argument, and returns the radius that encloses that mass under the DMB model. This displacement function represents the radial offset required to transform a DMO matter distribution into a DMB one.

The BCM proceeds as follows. We start with a DMO simulation with a halo catalog. For each halo above a the mass threshold $10^{13} M_{\odot}/h$\footnote{We chose a threshold of \( 10^{13} \, M_{\odot}/h \) for computational reasons: the halo mass function is dominated by low-mass halos, which would exponentially increase the pipeline's runtime. The impact of excluding these small halos is discussed in the results section.}, we identify all (dark matter) particles within a predetermined radius. Then, given the halo mass $M_{\rm 200c}$ and the radial separation of the particles from the halo, we evaluate the displacement $\Delta d(r)$. Next, we radially\footnote{\citet{anbajaganeMaplevelBaryonificationEfficient2024} found the impact of anisotropic (elliptical) offsets on the moments of the density field to be sub-percent (see their Figure 7).} offset the particles by $\Delta d(r)$ from the halo center. The displacements are generally positive on the scale of the halo radius, as the dark matter particles must be moved outwards to mimic the impact of extended gas distributions, and are generally negative on scales a hundredth of the halo radius, where the dark matter particles must coalesce to mimic the centrally peaked stellar distribution. The final particle distribution represents a density field that mimics the impact of baryonic processes. 

One can then vary the parameters controlling the mass profiles and thereby obtain different corrections to the simulated density field. The profiles specified by a given model can then also be used to predict other quantities, such as the integrated stellar/mass as a function of halo mass. The BCM approach can simultaneously fit variations in the matter power spectrum alongside other quantities such as the scaling relations of gas and star fractions with halo mass, the halo gas profiles, etc. \citep{schneiderQuantifyingBaryonEffects2019, giriEmulationBaryonicEffects2021,  To2024DMBCluster, Bigwood:2024:BaryonsWLkSZ}. 

Our discussions thus far describe the application of BCM to three-dimensional particle snapshots. However, for the purpose of weak lensing, it is computationally expensive to apply the BCM to each individual snapshot. To produce accurate lensing convergence maps, we require fine redshift resolution, e.g., $\mathcal{O}(100)$ snapshots, which increases the computational load of this task. The snapshot-based approach is also particularly expensive for the purpose of simulation-based inference, where of order $\mathcal{O}(10^4)$ simulations are needed to infer cosmological constraints. Therefore, we use the method outlined in \citet{anbajaganeMaplevelBaryonificationEfficient2024} which circumvents the computational cost by applying the BCM directly on 2D density shells on the full-sky.\footnote{An earlier version of this approach is also discussed in \citet{Fluri2022wCDMKIDS}, but does not contain a number of modeling elements introduced in \citet{anbajaganeMaplevelBaryonificationEfficient2024}.} This is done by rewriting the model in terms of \textit{projected} profiles and distances, rather than three-dimensional ones. Instead of particles, our units are the 2D pixels in the full-sky maps. 
For each pixel, we apply displacements from all halos --- where the given pixel is contained with an aperture of $20 R_{\rm 200c}$ around each selected halo --- and accumulate the contributions from all halos to compute the total displacement of the pixel. This extension allows for efficient modeling of higher-order correlations in weak lensing observables by making the BCM more scalable, and therefore makes the method more easily applicable to large datasets from current and upcoming cosmological surveys.

All baryonification in this work is carried out using the publicly available \textsc{BaryonForge}\footnote{\url{https://github.com/DhayaaAnbajagane/BaryonForge}} package \citep{anbajaganeMaplevelBaryonificationEfficient2024}. We direct readers to \citet{anbajaganeMaplevelBaryonificationEfficient2024} for more details on the model and method choices. 
For reference, generating 12 different baryonified simulations, each simulation consisting of 60 tomographic density field shells with $\nside = 2048$, takes approximately 4 hours on an AMD EPYC 7763 CPU node with 128 cores \comments{and 512 GB of memory. However, we only use a fraction of the memory as the memory footprint of the procedure is set by the density shells and the halo catalogs. For this work, the total memory footprint is less than 10 GB across all shells.}

The BCM parameters are listed in Table~\ref{tab:params_descrp}, along with their ranges and descriptions. As we will see, the three most important parameters -- the ones that can reproduce most of the variation across the hydrodynamic variants -- are:
\begin{itemize}
    \item $M_c$: a mass scale at which the characteristic slope of the gas profile is $\beta = 1.5$. Halos above/below this mass-scale have more/less cored gas profiles.
    \item $\thetaej$: a physical scale quantifying how extended the gas distribution is, expressed  in units of $R_{200c}$ of the halo. We will see that this ratio is of order 5, so the gas has quite an extended distribution.
    \item $\eta$: The power-law scaling of the stellar mass fraction with halo mass, $M_{\rm star}/M_{\rm 200c} \propto M_{\rm 200c}^\eta$.
\end{itemize}
\comments{The exact parameters we vary are primarily informed by the analysis of \citet{giriEmulationBaryonicEffects2021}. Compared to their work we also add the redshift scaling parameters $\nu_X$ (though, we will show below these are not required for accurate predictions) and $\eta$. The latter is added to give the model more flexibility in varying the relative fractions of the gas and star components in a halo.}

\begin{table}
    \centering
    \small
    \begin{tabular}{p{1.5cm} c p{10cm}}
       \toprule
       \textbf{Param}  & \textbf{Prior} & \textbf{Description} \\
       \midrule
       $\nu_{M_{\rm c}}$ & [-5, 5] & The redshift scaling of the gas mass scale. \\
       $\nu_{\theta_{\rm ej}}$ & [-5, 5] & The redshift scaling of the ejection radius. \\
       $\gamma$ & [0.1, 8] & The slope of the generalized NFW profile at $r \sim R_{\rm ej}$. \\
       $\log_{10}\theta_{\rm co}$ & [-4.0, -0.1] & The logarithm of the radial scale of the collapsed gas, $R_{\rm ej} = \theta_{\rm co}\Rtwohc$. \\
       $\log_{10}\eta_{\delta}$ & [-3.0, 0.0] & The power-law scaling for the central galaxy stellar fraction--halo mass relation. \\
       $\delta$ & [2, 20] & The slope of the generalized NFW profile at $r \gg R_{\rm ej}$. \\
       $\mu_{\beta}$ & [-3, 3] & The mass-dependent scaling of $\beta$, which controls the slope between $R_{\rm co} < r < R_{\rm ej}$. \\
       $\log_{10}\frac{M_{\rm c}}{\Msun}$ & [12.0, 16.0] & The logarithm of the mass-scale where the characteristic gas slope transitions. \\
       $\log_{10}\eta$ & [-3.0, 0.0] & The logarithm of the power-law scaling of stellar fraction--halo mass relation. \\
       $\theta_{\rm ej}$ & [0.8, 20] & The radial scale of the ejected gas, $R_{\rm ej} = \theta_{\rm ej}\Rtwohc$. \\
    \bottomrule
    \end{tabular}
    \caption{The input parameters to the profiles of the BCM. For each parameter, we list its prior range and a description of the parameter's physical meaning. See \citet{anbajaganeMaplevelBaryonificationEfficient2024} for the specific equations that these parameters govern. The prior ranges are similar to those in \cite{anbajaganeMaplevelBaryonificationEfficient2024,Pandey2024godmax} and are used in \Sec\ref{sec:emulator} to construct samples of the baryon-corrected maps and to train the BCM emulator.}
    \label{tab:params_descrp}
\end{table}

%% file: 3_sim.tex
\section{Simulations}\label{sec:sim}
\subsection{\flamingo{} simulations}

Our analysis relies on the \flamingo{} suite of cosmological hydrodynamical simulations, all of which share the initial conditions of a DMO simulation. We provide a brief summary here; for more details, see \cite{Schaye2023,Kugel2023}. 
The simulations were run using the SPHENIX smoothed particle hydrodynamics implementation \citep{Borrow2022} within SWIFT \citep{Schaller2024}. \flamingo{} includes radiative cooling and heating \citep{Ploeckinger2020}, star formation \citep{Schaye2008}, time-dependent stellar mass loss \citep{Wiersma2009}, massive neutrinos \citep{Elbers2021}, and kinetic supernova and stellar feedback \citep{DallaVecchia2008,Chaikin2023}. The simulations are calibrated to match the galaxy stellar mass function at $z = 0$ and the gas fraction of low-$z$ clusters and groups of clusters inferred from X-Ray and weak lensing observations \citep{Kugel2023}. \comments{Halos are identified using a three-dimensional friends-of-friends (FoF) algorithm and we utilize the corresponding FoF masses of these halos.}\footnote{\comments{While it is more correct to use spherical overdensity (SO) masses for the baryonification method, we did not have the relevant SO masses for every simulation of interest during the start of this project when the baryonified maps were generated. We have thus used FoF masses instead, which were available for all simulations. While the use of FoF masses over SO masses will lead to some difference (and cause small shifts in the inferred parameters), it does not affect our ability to quantify the flexibility of the baryonification model for the statistics of interest.}}

We focus on AGN feedback, implemented in two forms: thermal \citep{Booth2009}, where particles in the AGN vicinity are heated, and kinetic jet-like, where particles are given momentum kicks along the black hole (BH) spin axis \citep{Husko2022}. We consider nine variations in AGN feedback strength from the models available in the \flamingo{} simulations, all in $1~\mathrm{Gpc}^3$ volumes:

\begin{itemize}
\item {{\bf Fiducial}: This model features thermal, AGN feedback without jets. Outflows follow paths of least resistance, forming buoyant high-entropy gas bubbles in clusters. Energy is injected into the nearest gas particle, reducing feedback at large distances compared to mass-weighted schemes. The AGN heating temperature, $\Delta T_{\text{AGN}}$, is calibrated to match observations such as the cluster gas fractions}. This model is labeled Fiducial (L1\_m9) per \flamingo{} convention.

\item Thermal AGN – cluster gas and galaxy stellar mass function variations: We include {\bf 6 variations} of the fiducial model where AGN feedback remains thermal, but the simulations feature different cluster gas fractions ($\Delta f_{\text{gas}}$) and/or galaxy stellar mass functions ($\Delta \Phi_*$), shifted by a specified number of observational standard deviations ($\sigma$). These variations lead to different AGN feedback strengths. The models are labeled fgas$+2\sigma$, fgas$-2\sigma$, fgas$-4\sigma$, fgas$-8\sigma$, M$-\sigma$, and M$-\sigma$\_fgas$-4\sigma$.

\item Jet-like AGN Feedback: The {\bf two additional models}, Jet and Jet\_fgas$-4\sigma$, replace the Fiducial thermal AGN feedback with kinetic, jet-like feedback. Compared to the Fiducial model, Jet has slightly weaker stellar feedback, while Jet\_fgas$-4\sigma$ features higher jet velocity.
\end{itemize}

We apply the baryonification procedure described in \Sec\ref{sec:bary} to the full-sky spherical overdensity shells of the DMO simulation to generate baryonified realizations. \comments{See Appendix A of \citet{Schaye2023} for details on how the lightcone is generated.} \comments{Our procedure for making lensing maps is detailed below in Section \ref{sec:convergencemaps}.} All simulations -- both hydrodynamical and baryonified -- adopt the best-fit $\Lambda$CDM cosmology from DES Y3 “3×2pt + All Ext.” \citep{Abbott2022}\comments{, which is $h = 0.681, \Omega_{\rm m} = 0.306, \sigma_8 = 0.807$. For the remaining parameters, see Table 4 in \citet{Schaye2023}.}

Finally, to investigate the cosmology dependence of baryonic feedback in \Sec\ref{sec:result}, we also consider simulations run at the “LS8” cosmology \citep{amonConsistentLensingClustering2022,Schaye2023}, which has a lower amplitude of the power spectrum: $S_8 = 0.766$ compared to $0.815$ for the fiducial cosmology. The simulations include DMO (LS8\_DMO), the fiducial AGN (LS8), and the strong AGN (LS8\_fgas-8$\sigma$) variants. We did not apply the BCM to the LS8\_DMO simulation and used these three simulations exclusively for the cosmology dependence analysis in \Sec\ref{sec:result}.

\subsection{DES-like weak lensing convergence maps}\label{sec:convergencemaps}
\begin{figure}
    \centering
    \includegraphics[width=0.45\linewidth]{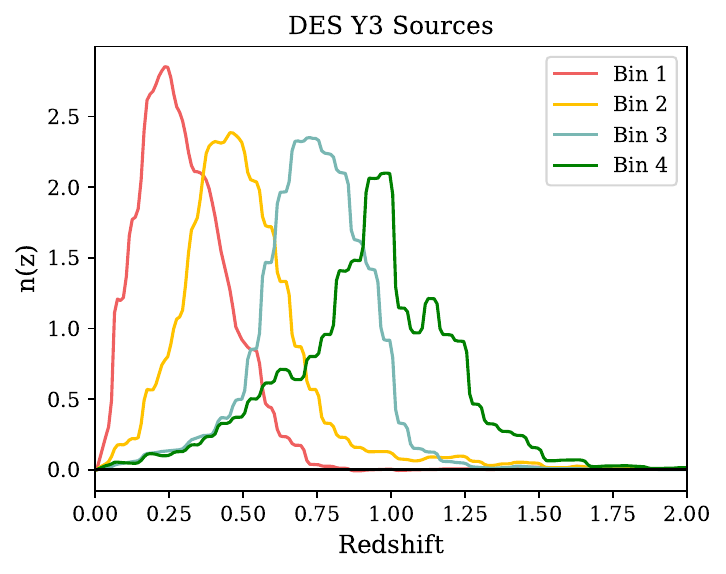}
    \includegraphics[width=0.45\linewidth]{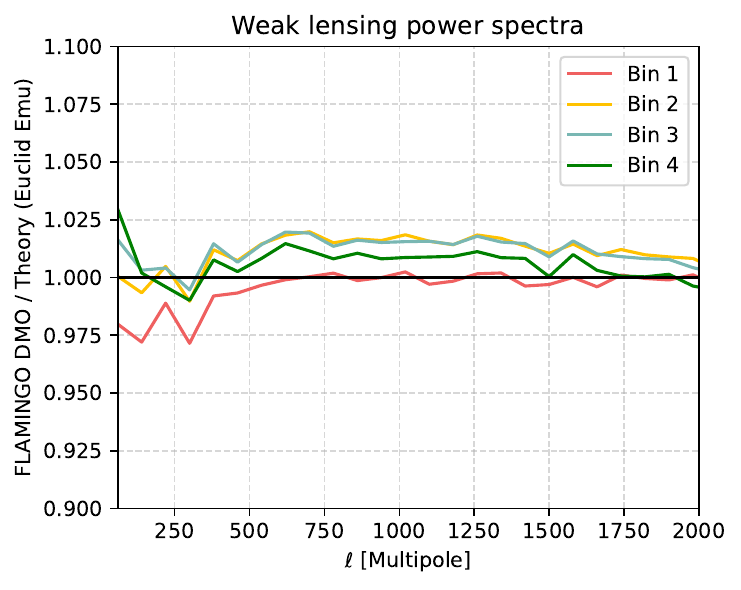}

    \caption{\textit{Right:} redshift distribution of the four tomographic bins of the DES Y3 wak lensing sample considered here.
    \textit{Left:} Power spectra of the 4 noiseless, full-sky convergence maps produced from the \flamingo{} DMO simulation, using the DES Y3 redshift distributions, compared to predictions from EuclidEmu \citep{Knabenhans2019}. }
    \label{fig:flamingo_cls}
\end{figure}

We post-process the \flamingo{} simulations to generate weak lensing convergence maps, $\kappa$. Each \flamingo{} simulation provides lightcone outputs in the form of HEALPix maps \citep{GORSKI2005} as a function of redshift. The native HEALPix resolution of $\nside = 16384$ is degraded to $\nside = 2048$ for this project (corresponding to a pixel scale of approximately $1.4$ arcminutes), with shell thickness $\Delta z = 0.05$. We produce noiseless, full-sky convergence maps assuming the DES Y3 weak lensing redshift distributions in Fig.\ref{fig:flamingo_cls} from \citealt{y3-sourcewz,Myles2021PhotoZ}. \comments{Maps are made assuming the Born approximation.}\footnote{\comments{\citet{Petri2017Born} show that post-born corrections, estimated via a full ray-tracing algorithm, result in of $\mathcal{O}(10\%)$ corrections to statistics on small-scales. This is a statement on the absolute accuracy of a simulation compared to the observed lensing process. In this work we are focused on relative differences between simulations, where the relevance of this effect is greatly suppressed. Note also that \citet[][see their Figure 1]{Broxterman:2025:RayTraced} show the Born approximation produces a consistent (within 1\%) result for the angular power spectrum of the matter field for scales up to $\ell < 10^4$}} Mimicking the DES Y3 tomographic sample, we generate 4 convergence maps by summing the shells weighted by the appropriate redshift distribution, enabling us to study baryonic effects as a function of redshift. Fig.\ref{fig:flamingo_cls} shows the resulting 2D DMO power spectra in the four bins compared to theoretical predictions from the Euclid Emulator\citep{Knabenhans2019}, with agreement at the 1–2\% level, consistent with the emulator’s accuracy. \comments{We do not add the signal from the intrinsic alignments (IA) of galaxies to our maps, and the coupling of IA and baryonic feedback is minimal \citep{Tenneti:2017:IA, Soussana:2020:AGNIA}.}

Since our focus is on the relative impact of baryonic feedback compared to the DMO simulation, we do not forward-model the DES Y3 noise or mask, and work exclusively with full-sky, noiseless convergence maps. In total, we use three sets of tomographic maps:
\begin{itemize}
\item $\kappadmo$: one set from the DMO simulation,
\item $\kappahydro$: 9 sets from the nine \flamingo{} feedback variants,
\item $\kappabary$: $\sim 5,000$ sets, generated by baryonifying $\kappadmo$ following the BCM prescription in \Sec\ref{sec:bary}, discussed further in \Sec\ref{sec:emulator}.  
\end{itemize}

%% file: 4_stats.tex
\section{Statistics}\label{sec:stat}
\begin{figure}
    \centering
    \includegraphics[width=0.95\textwidth]{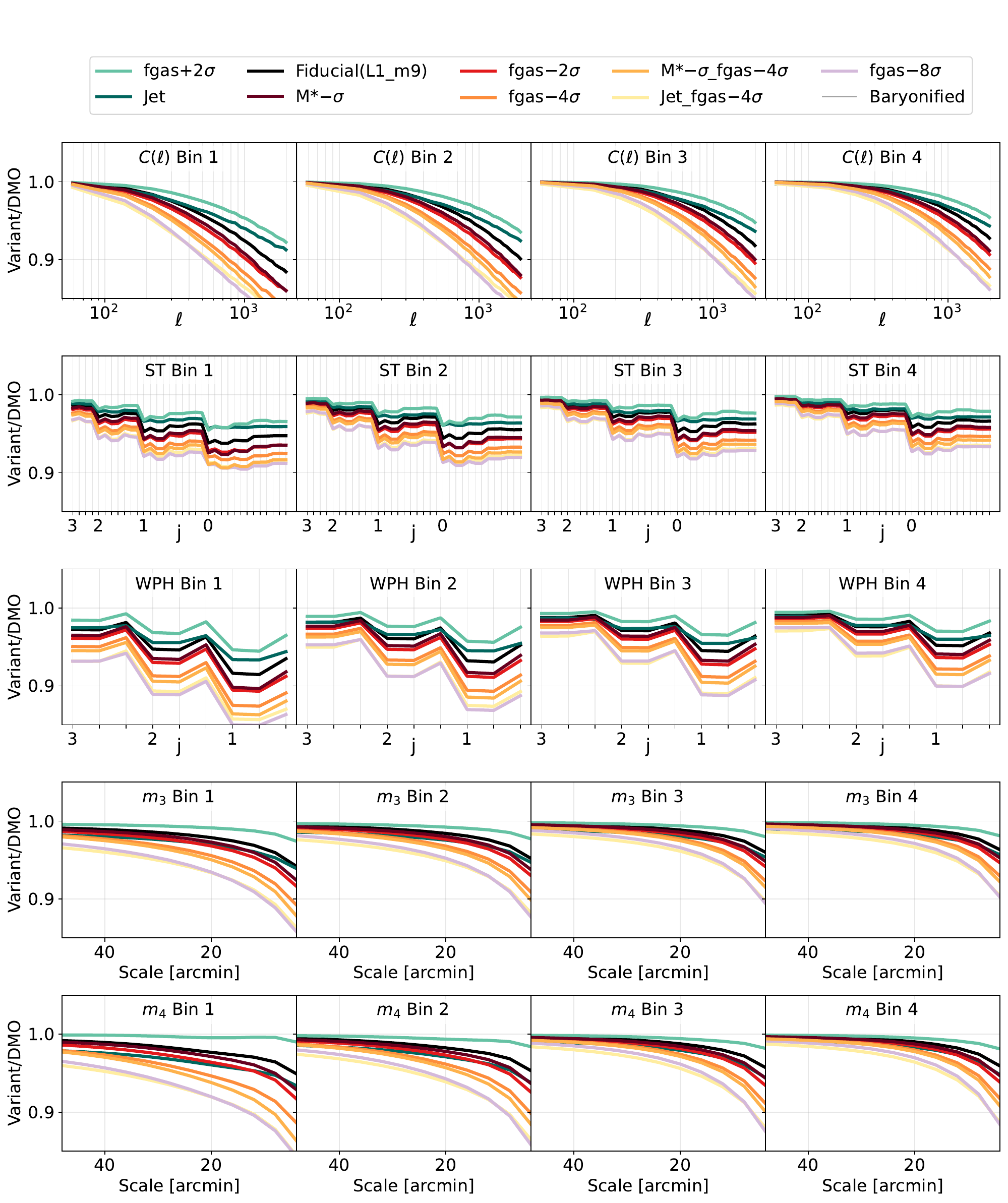}
    \caption{Summary statistics of the \flamingo{} $\kappa$ maps (\Sec\ref{sec:sim}). Rows correspond to different statistics, and columns to redshift bins (modeled after DES Y3 data). The y-axis shows the \flamingo{} statistics relative to the statistics of the DMO simulation of the same initial condition. On the x-axis, spatial scales decreases to the right. ST and WPH are ordered by the wavelet size ($j$) as a proxy for their characteristic scale. In \flamingo{}, baryon feedback physics suppresses all statistics on all scales, and is the strongest in low redshift bins.}
    \label{fig:data}
\end{figure}

We now describe the theory, measurement, and scale cuts of the statistics used here: the power spectra ($C(\ell)$), the scattering transform coefficients (ST), the wavelet phase harmonic coefficients (WPH), and the third- ($m_3$) and the fourth-moments ($m_4$) of the $\kappa$ maps. As an example, the statistics of $\kappahydro$ are shown in \Fig\ref{fig:data} in colored lines. We also summarize all the statistics in \Tab\ref{tab:stats_summary} at the end of the section. \comments{All statistics are computed on maps of $N_{\rm side} = 1024$, corresponding to a pixel scale of $3.44 \arcmin$. At $z = 0.5$, which is where the lensing kernel peaks \citep[e.g.,][]{Secco2022}, this corresponds to a comoving distance of $d = 1.9 \Mpc$. This is computed assuming the cosmology of the Flamingo simulation used in this work.}

\subsection{Power Spectrum}\label{sec:stat_ps}
The power spectrum of a $\kappa$ map is defined as
\begin{equation}
\label{eqn:cl}
    C(\ell) = \frac1{2l+1}\,\sum_m \left| \hat{\kappa}_{\ell m} \right|^2 \,,
\end{equation}
where $\hat{\kappa}_{\ell m}$ is the spherical harmonic transform of $\kappa$ that is downgraded to $\nside=1024$. We consider the auto-power spectra of the 4 DES redshift bins, linearly binned in $\ell \in [20,2020]$.

A main objective is to validate BCM's weak lensing statistics to spatial scales smaller than the previous DMO analyses. For example, the DES Y3 weak lensing simulation-based inference with HOS was limited to $\ellcut = 1000$ \citep{gattiDarkEnergySurvey2023, gattiDarkEnergySurvey2024}, because the analysis assumed a DMO model and could not model smaller scales contaminated by baryonic feedback. Here, we aim to validate the BCM to $\ellcut = 2000$. 
We do not attempt to model $\ell > 2,000$ due to 1) increased systematic effects and a low signal-to-noise ratio in real data, and 2) the computational challenge of producing enough simulations with low shot noise at high $\ell$ in the context of simulation-based inference. 

\subsection{Scattering Transform Coefficients (ST)}\label{sec:stat_st}
Wavelets are localized oscillatory functions that act as band-pass filters in both real and Fourier space. Unlike Fourier analysis which is a global transformation, wavelets provide spatial localization while simultaneously capturing frequency information. A wavelet family consists of wavelets of different scales and orientations, and can be used to extract multi-scale patterns in images while preserving directional information \citep{mallatGroupInvariantScattering2012}.

Given a template wavelet $\psi(\bfr)$,\footnote{For scattering transform, we use the Morlet wavelet under the convention of \cite{chengNewApproachObservational2020}.} we can generate new wavelets of different sizes and orientations via
\begin{equation}
    \psi_{j,l}(\bfr) \propto \psi \Big(\frac{1}{2^j}R_{l\pi/L}^{-1} \bfr\Big)\,,
    \quad j \in \{\Z \mid 0 \leq j < J\}\,,
    \quad l \in \{\Z \mid 0 \leq l < L\}\,.
\end{equation}
Here $j$ and $l$ label the scale and the orientation of the derived wavelets. Each wavelet covers $\sim 2^j$ image pixels, and larger $j$ corresponds to larger scales. The operator $R^{-1}$ in the argument is the rotation matrix with the angle of rotation given by the subscript, with the wavelets covering $L$ directions uniformly distributed over $[0,\pi)$.

The scattering transformed $\kappa$ maps are constructed by iteratively applying the wavelet convolution and the modulus operator \citep{mallatGroupInvariantScattering2012, chengNewApproachObservational2020}. The first and second-order scattering transformed $\kappa$ maps are
\begin{align}
    \kappa^{\mathrm{ST},(1)}_{j,l} &= \left| \kappa \star \psi_{j,l} \right| \,, \\
    \kappa^{\mathrm{ST} \,,(2)}_{j,l,j',l_\Delta } &= \left| \,\left| \kappa \star \psi_{j,l} \right| \star \psi_{j',l - l_\Delta } \right| \,,
\end{align}
where $\star$ denotes convolution. The modulus preserves the first-order field statistics while introducing the necessary non-linearity to extract information beyond the mean without the instability to outliers that higher-order moments would cause. The higher-order convolution extracts information about spatial distributions $\kappa^{\mathrm{ST} \,,(1)}$, and captures HOS by measuring the ``clustering of clustered structures,'' with each order $n$ potentially containing information up to $2^n$-point functions \citep{carronINCOMPLETENESSMOMENTCORRELATION2011, chengNewApproachObservational2020}. For the second-order field, we typically require $j \leq j'$ to ensure that the second wavelet extracts information at scales equal to or larger than the first. Since $\kappa$ is a homogeneous and isotropic field, we compress $\kappa^{\mathrm{ST}}$ through spatial and angular averaging \citep{allysRWSTComprehensiveStatistical2019, chengNewApproachObservational2020}. More specifically, the ST coefficients are given by
\begin{align}
\label{eqn:st1}
    s^{(1)}_{j} &= \langle \kappa^{\mathrm{ST},(1)}_{j,l} \rangle_{p, l} \,, \\
\label{eqn:st2}
    s^{(2)}_{j,j',l_\Delta} &= \langle \kappa^{\mathrm{ST} \,,(2)}_{j,l,j',l_\Delta} \rangle_{p, l},
\end{align}
where $\langle \cdot \rangle$ denotes the expectation value, and the subscripts indicate averaging over pixels ($p$) and orientations ($l$). 
These coefficients compactly and robustly summarize the information in the hierarchical scales of the clustering signal beyond the power spectrum.

To measure ST, we downgrade the $\kappa$ maps to $\nside=1024$, and divide them into $\nside=8$ HEALPix pixels, each gnomonically projected to a $256 \times 256$ Cartesian patch with $3.44\arcmin$ resolution. Each patch is centered on an $\nside=8$ pixel with surrounding HEALPix pixels masked to prevent double counting. We measure ST on $1/8$ of the sky using the \texttt{scattering\_transform} package\footnote{\texttt{github.com/SihaoCheng/scattering\_transform}} with parameters $J = 4$ and $L = 3$, leading to 4 $s^{(1)}$'s and 30 $s^{(2)}$'s per redshift bin. The final ST is calculated independently for each tomographic bin and averaged across all Cartesian patches. 

Since ST does not have a sharp scale cutoff like $C(\ell)$, we empirically determine the characteristic scale of each ST, and perform consistent scale cut in \App\ref{appx:scale}. All 34 ST coefficients per redshift bin are found to predominantly probe scales below $\ellcut = 2000$.

\subsection{Wavelet Phase Harmonics Coefficients (WPH)}\label{sec:stat_wph}
WPH is another class of wavelet-based summary statistics.
Again, we use $\psi_{j,l}(\bfr)$ to denote wavelets of different sizes ($j$) and orientations ($l$), but with a different waveform -- the bump-steerable wavelets \citep{mallatPhaseHarmonicCorrelations2019}.
WPH statistics are obtained by first convolving $\kappa$ with $\psi_{j,l}(\bfr)$, then taking its phase harmonics, and finally computing the moments of the resulting fields \citep{mallatPhaseHarmonicCorrelations2019, allysNewInterpretableStatistics2020}. Since $\kappa$ is isotropic, we also average over the orientations of the wavelets. The equations for the first- and second-order WPH reduce to
\begin{align}
\label{eqn:wph1}
    \wph{1}{1}_{j} &= \langle \mathrm{Cov}\bigl(\kappa \star \psi_{j, l},\, \kappa \star \psi_{j, l}\bigr) \rangle_l \,, \\
\label{eqn:wph2}
    \wph{0}{0}_{j} &= \langle \mathrm{Cov}\bigl(|\kappa \star \psi_{j, l}|,\,|\kappa \star \psi_{j, l}|\bigr) \rangle_l \,, \\
\label{eqn:wph3}
    \wph{0}{1}_{j} &= \langle \mathrm{Cov}\bigl(|\kappa \star \psi_{j, l}|,\,\kappa \star \psi_{j, l}\bigr) \rangle_l \,.
\end{align}
$\wph{0}{0}$ and $\wph{0}{1}$ contain both Gaussian and non-Gaussian information due the nonlinearity of the phase harmonics (in this case, the modulus) operation. $\wph{1}{1}$ contains only Gaussian information, albeit with a different smoothing scheme compared to the power spectrum.

The measurement of WPH is similar to that of ST\comments{, on gnomonically projected patches of $3.44\arcmin$ resolution.} We again work with Cartesian patches that are projected from the $\kappa$ maps. On each patch, we compute WPH using the \texttt{PyWPH} package\footnote{\texttt{github.com/bregaldo/pywph}}. We consider wavelets with $J = 4$ and $L = 3$, leading to 4 of each $\wph{1}{1}$, $\wph{0}{0}$, and $\wph{0}{1}$ per redshift bin. We then average the WPH from each patch across $1/8$ of the sky to reduce cosmic variance. 

The scale cut for WPH is empirically determined in \App\ref{appx:scale} and is consistent with the $\ellcut=2000$ scale cut for the power spectra. The three $j=0$ WPH are removed from subsequent analyses.

\subsection{Higher Order Moments}\label{sec:stat_mom}
The third and fourth moments of the convergence field are defined as
\begin{align}
\label{eqn:m3}
    m_3(\theta) &= \bigl\langle \bigl(\kappa \star G(\theta)\bigr)^3 \bigr\rangle_p \,, \\
\label{eqn:m4}
    m_4(\theta) &= \bigl\langle \bigl(\kappa \star G(\theta)\bigr)^4 \bigr\rangle_p \,,
\end{align}
where $G(\theta)$ is a Gaussian kernel of width $\theta$ and $p$ denotes pixels. $m_3$ measures skewness of $\kappa$. It contains only non-Gaussian information and is strictly zero for a Gaussian random field. $m_4$ measures the kurtosis of $\kappa$ and has both a Gaussian (derivable from the power spectrum) and a non-Gaussian contribution following Wick's theorem. We set the minimum scale at $4\arcmin$ and use $\theta \in [4\arcmin, 48 \arcmin]$ in increments of $4 \arcmin$. This is consistent with the scale cuts of other statistics since a $4 \arcmin$ Gaussian-smoothed $\kappa$ has negligible power beyond $\ell = 2000$. 

The modeling and the real data measurement of $m_3$ and $m_4$ are more sensitive to noise than two-point statistics. This is because 1) by the central limit theorem, for a sample with size $N$, while $\langle \kappa \rangle$ has an error $\propto N^{-1/2}$, for $\langle \kappa^n \rangle$, the effective standard error scales more like $\alpha N^{-1/2}$ where $\alpha$ grows superlinearly with $n$, and 2) the tail of outliers of the distribution contributes disproportionally, amplifying the fluctuations. Perturbation theory provides a similar perspective, where empirically $\bigl\langle \kappa^n \bigr\rangle \propto \bigl\langle \kappa^2 \bigr\rangle^{n-1}$ \citep{bernardeauLargeScaleStructureUniverse2002a}. Therefore an $x$ uncertainty in $\langle \kappa^2 \rangle$ propagates to $(n-1) x$ in $\langle \kappa^n \rangle$.

\subsection{The statistics of \flamingo{} and baryonified convergence maps}\label{sec:stat_flamingobary}
In total, we consider 87 (348) data points per redshift bin (and in total), as summarized in \Tab\ref{tab:stats_summary}. \comments{Throughout this work, we consider summaries of the auto-correlation of the four tomographic bins, but do not include their cross-correlations for the sake of reducing the size of the data vector. We still test the ability of the baryonification model to fit all four tomographic bins, and are therefore adequately probing the model's redshift evolution (as it pertains to weak lensing analyses).} The statistics for $\kappahydro$, relative to those of $\kappadmo$, are shown in \Fig\ref{fig:data}. In all \flamingo{} feedback variants, baryons suppress all statistics at all scales. The suppression is stronger at smaller scales and lower redshifts, consistent with previous works \citep[e.g.,][]{vandaalenEffectsGalaxyFormation2011,Chisari2018BaryonsPk,huangModelingBaryonicPhysics2019,amonNonlinearSolution82022,grandonImpactBaryonicFeedback2024a,schallerAnalyticRedshiftindependentFormulation2025}. Relative to the fiducial model, the fgas$+2\sigma$ variant shows weaker baryonic signatures, whereas the fgas$-2\sigma$ and Jet\_fgas$-4\sigma$ variants induce stronger suppression in the statistics.
\begin{table}[ht]
    \centering
    \small
    \begin{tabular}{llccc}
    \toprule
    \multicolumn{2}{c}{\textbf{Statistics}} & \textbf{Equation} & \textbf{N per bin} & \textbf{Notes} \\
    \midrule
    \multirow{2}{*}{ST} 
        & $s^{(1)}_{j}$ & \Eq\ref{eqn:st1} & 4 & $0 \leq j < 4$\\
        \addlinespace
        & $s^{(2)}_{j,j',l_\Delta}$ & \Eq\ref{eqn:st2} & 30 & $0 \leq j \leq j' < 4$, $0 \leq l_\Delta < 3$ \\ 
    \addlinespace
    \midrule
    \addlinespace
    \multirow{3}{*}{WPH} 
        & $S^{(1,1)}$ & \Eq\ref{eqn:wph1} & 3 & $1 \leq j < 4$\\
        & $S^{(0,0)}$ & \Eq\ref{eqn:wph2} & 3 & $1 \leq j < 4$\\
        & $S^{(0,1)}$ & \Eq\ref{eqn:wph3} & 3 & $1 \leq j < 4$\\
    \addlinespace
    \midrule
    $C(\ell)$ & & \Eq\ref{eqn:cl} & 20 & linear bin, $\ell \in [20,2020]$\\
    \midrule
    $m_3(\theta)$ & & \Eq\ref{eqn:m3} & 12 & smoothing scale $\theta \in [4\arcmin, 48 \arcmin]$\\
    \midrule
    $m_4(\theta)$ & & \Eq\ref{eqn:m4} & 12 & smoothing scale $\theta \in [4\arcmin, 48 \arcmin]$ \\
    \bottomrule
    \end{tabular}
    \caption{Summary of statistics used in the analysis, restricted to scales with $\ell \lesssim 2000$.}
    \label{tab:stats_summary}
\end{table}

\begin{figure}
    \centering
    \includegraphics[width=1\linewidth]{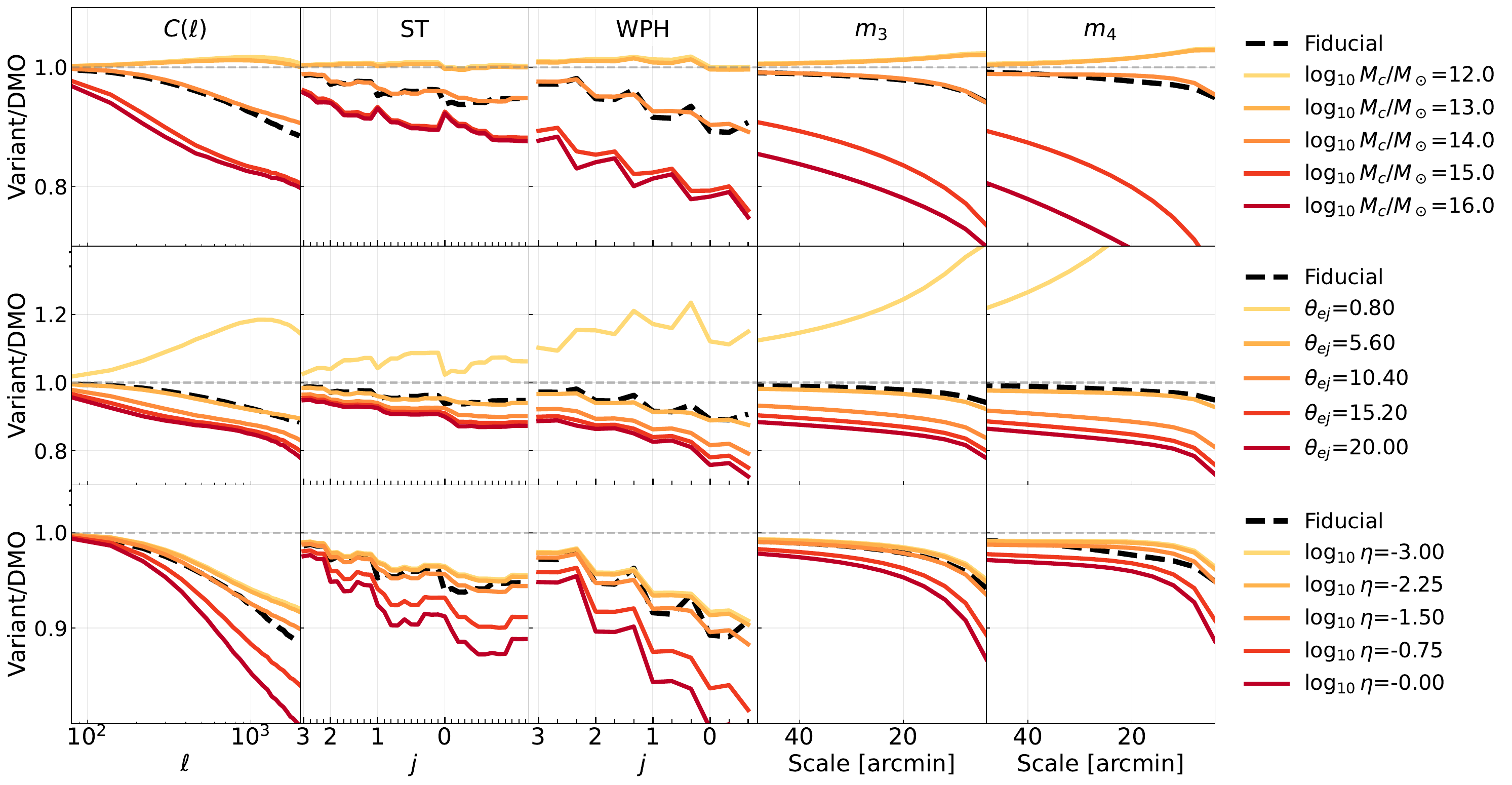}
    \caption{Summary statistics of the baryonified (colored) and the \flamingo{} fiducial variant's (black, dashed) convergence map, focusing on the first redshift bin. For the baryonified maps, a single BCM parameter is varied at a time. Rows correspond to different BCM parameters, and columns to different statistics. The y-axis shows the statistics of both the baryonified and \flamingo{} maps, relative to those of the DMO simulation with the same initial condition. On the x-axis, spatial scale decreases to the right. ST and WPH statistics are ordered by wavelet size ($j$), used here as a proxy for characteristic scale. The BCM flexibly generates a range of baryonic signatures in the statistics, clustering around the \flamingo{} results. The variation of other parameters in \Tab\ref{tab:params_descrp} is shown in \Fig\ref{fig:variation_7param} in \App\ref{appx:sensitivity}.}
    \label{fig:variation_3param}
\end{figure}

To build intuition about how each BCM parameter affects the statistics, we vary each parameter independently over its prior range defined in \Tab\ref{tab:params_descrp}, and plot the resulting statistics of $\kappabary$ relative to those of $\kappadmo$ in Fig. \ref{fig:variation_3param}. Our final results in \Sec\ref{sec:result} suggest that the three most important parameters to match and capture variations among different FLAMINGO feedback models are $M_c$, $\theta_{\rm ej}$, and $\eta$, which we highlight in \Fig\ref{fig:variation_3param}. The effects of the remaining parameters are shown in \Fig\ref{fig:variation_7param} in \App\ref{appx:sensitivity}. We find that the prior ranges induce a wide range of changes in the statistics—including both suppression and, in some cases, enhancement—relative to the fiducial variant’s measurement. 

%% file: 5_method.tex
\section{Method}\label{sec:method}
We want to demonstrate that the BCM can flexibly and accurately reproduce the wide range of baryonic feedback models in \flamingo{}, in terms of two-point and HOS and investigate what are the baryon parameter values that best fit the \flamingo{} simulations and how they vary as a function of hydrodynamical feedback strengths.

In principle, we could sample the best-fit baryon parameters in one step using a Markov Chain Monte Carlo (MCMC) -- given the baryon parameters, we could generate baryonified $\kappa$ maps using the DMO simulation, measure their statistics, and compare them to those from \flamingo{}. In practice, generating new baryonified maps and computing their statistics at each MCMC step is too time-consuming. Therefore, we proceed in three steps:
\begin{enumerate}
    \item Build an emulator that predicts the statistics for the given baryon parameters
    \item Use the emulator in an MCMC on each of the \flamingo{} simulations to find the best-fit baryon parameters
    \item Once the MCMC chains have converged, we generate baryonified $\kappa$ maps at the MCMC best fit, measure the suppression of the statistics, and use them for our final considerations.
\end{enumerate}
Therefore, we use the emulator as a fast and efficient tool to refine our search in parameter space, rather than brute-forcing the problem by baryonifying the maps within the MCMC, which would be computationally infeasible. This means that possible discrepancies between the suppression of the statistics in the FLAMINGO simulations and the one we obtain at the end of our three-step procedure can be attributed to two factors: (1) the accuracy of the emulator in reproducing the statistics compared to the full baryonification process, and (2) the intrinsic limitations of the baryonification method itself, even assuming a perfect emulator.

\subsection{Emulator}\label{sec:emulator}
We will refer to the collection of all summary statistics as $\stat$.
To accelerate the MCMC inference, we construct an emulator $r$, which predicts the ratio of the BCM-corrected summary statistics to its DMO counterpart given baryon parameters $\bary$, i.e.,
\begin{equation}
r(\bary) = \statbary(\bary) / \stat_\mathbf{dmo} ,.
\end{equation}
The numerator $\statbary(\bary)$ is computed by first baryonifying $\kappadmo$ using the BCM (\Sec\ref{sec:bary}) and then performing two-point and HOS measurements (\Sec\ref{sec:stat}). Since $\kappabary$ and $\kappadmo$ share the same initial condition, the ratio eliminates shot noise and cosmic variance.

\begin{figure}
    \centering
    \includegraphics[width=1\linewidth]{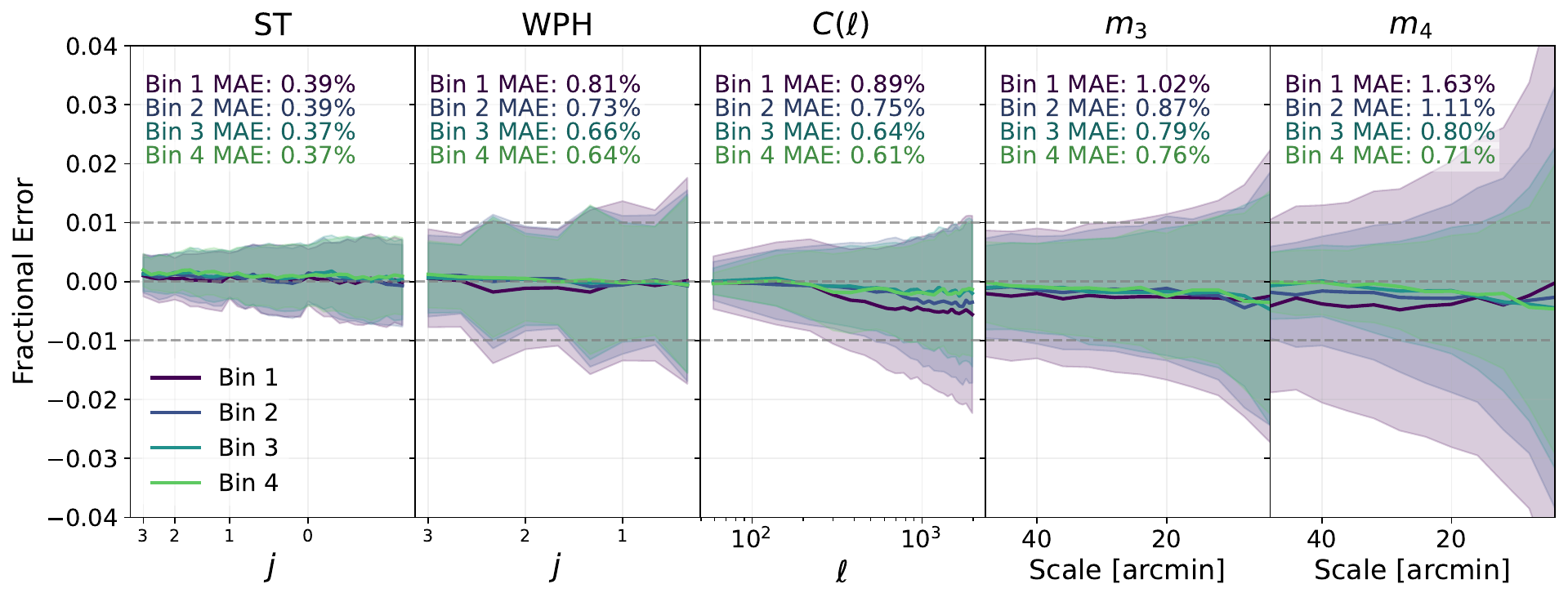}
    \caption{The emulator mean relative error, grouped by statistics and by redshift bins. The mean absolute (relative) error (MAE) is shown in text. The error is evaluated by averaging across the test set across the baryon parameter space. The emulator achieves sub-percent accuracy for all statistics besides the low redshift moments. The higher-order moment statistics are intrinsically nosier since it involves multiplications of random variables. We remind the reader that the emulator is used to efficiently explore the BCM parameter space when comparing to FLAMINGO statistics. However, the final assessment of how well the BCM reproduces a given FLAMINGO statistic is based on running the full baryonification pipeline at the emulator-derived best-fit parameters. }
    \label{fig:emulator}
\end{figure}

The training set includes $5,468$ $\kappabary$ maps sampled in the ten-dimensional baryon parameter space. The prior ranges for the parameters are detailed in \Tab\ref{tab:params_descrp} and are similar to those used in \citet{schneiderQuantifyingBaryonEffects2019,anbajaganeMaplevelBaryonificationEfficient2024}. In \Figs\ref{fig:variation_3param} and \ref{fig:variation_7param}, we have shown how varying a single BCM parameter can produce suppression similar to those observed in the fiducial model. We also explicitly verify that, across the training samples, the baryonified maps exhibit baryonic signatures spanning the full range of variations seen in the different \flamingo{} feedback variants. For each statistic, we construct an ensemble of 10 neural networks via cross-validation. We use the mean and 1$\sigma$ scatter of the ensemble predictions as the point estimate and uncertainty of $r(\bary)$. The emulator accuracy is shown in \Fig\ref{fig:emulator} as a function of scale and redshift, and generally decreases at smaller scales and lower redshifts, where baryonic feedback is more nonlinear and varied. The emulator achieves a mean absolute error below 1\% for all statistics except the low-redshift moments, which are intrinsically noisier, as discussed in \Sec\ref{sec:stat_mom}. The emulator is implemented in \jax{} and supports automatic differentiation \citep{jax2018github}. Details of the training set and emulator are provided in \App\ref{appx:emulator}.

\subsection{Sampling}
For each \flamingo{} feedback variant with statistics $\stat_{\mathbf{hydro}}$, we seek the best-fit baryon parameters by sampling the posterior $\mathcal{P}$
\begin{equation}
    \mathcal{P}(\bary\mid\stat_{\mathbf{hydro}}) = \distgauss\left(
    \frac{\stat_{\mathbf{hydro}}}{\stat_\mathbf{dmo}} - r(\bary), \sigma^2+\epsilon(\bary)^2
    \right)
\end{equation}
where $\distgauss$ is a Gaussian distribution parametrized by mean and variance, $\sigma$ is a constant uncertainty per statistic and $\epsilon$ is the emulator uncertainty. We assume a flat prior on $\bary$. As discussed above, the simulations are noiseless and the ratio statistics suppress the cosmic variance and shot noise. Therefore, in theory, the emulator uncertainty $\epsilon$ is the only source of error in $r(\bary)$. However, since $\epsilon$ is a function of $\bary$, the parameter-dependent covariance makes the sampling very difficult in high-dimensional space. Therefore, we introduce a constant noise $\sigma$ to promote the smoothness of the posterior surface to facilitate sampling. In particular, $\sigma$ does not reflect real observational uncertainty. We find $\sigma = 0.04$ for ST, WPH, and $C(\ell)$ and $\sigma = 0.08$ for $m_3$ and $m_4$ make the chains converge robustly. Although we are putting less weight on the moments, we shall see that the final best-fit baryon parameters reproduce all statistics equally well. We leverage the gradient information provided by the emulator and use the \numpyro{} implementation \citep{binghamPyroDeepUniversal2018, phanComposableEffectsFlexible2019} of the Hamiltonian Monte Carlo algorithm for efficient sampling \citep{nealMCMCUsingHamiltonian2011,betancourtConceptualIntroductionHamiltonian2018}.

%% file: 6_results.tex
\begin{figure}
    \centering
    \includegraphics[width=1\textwidth]{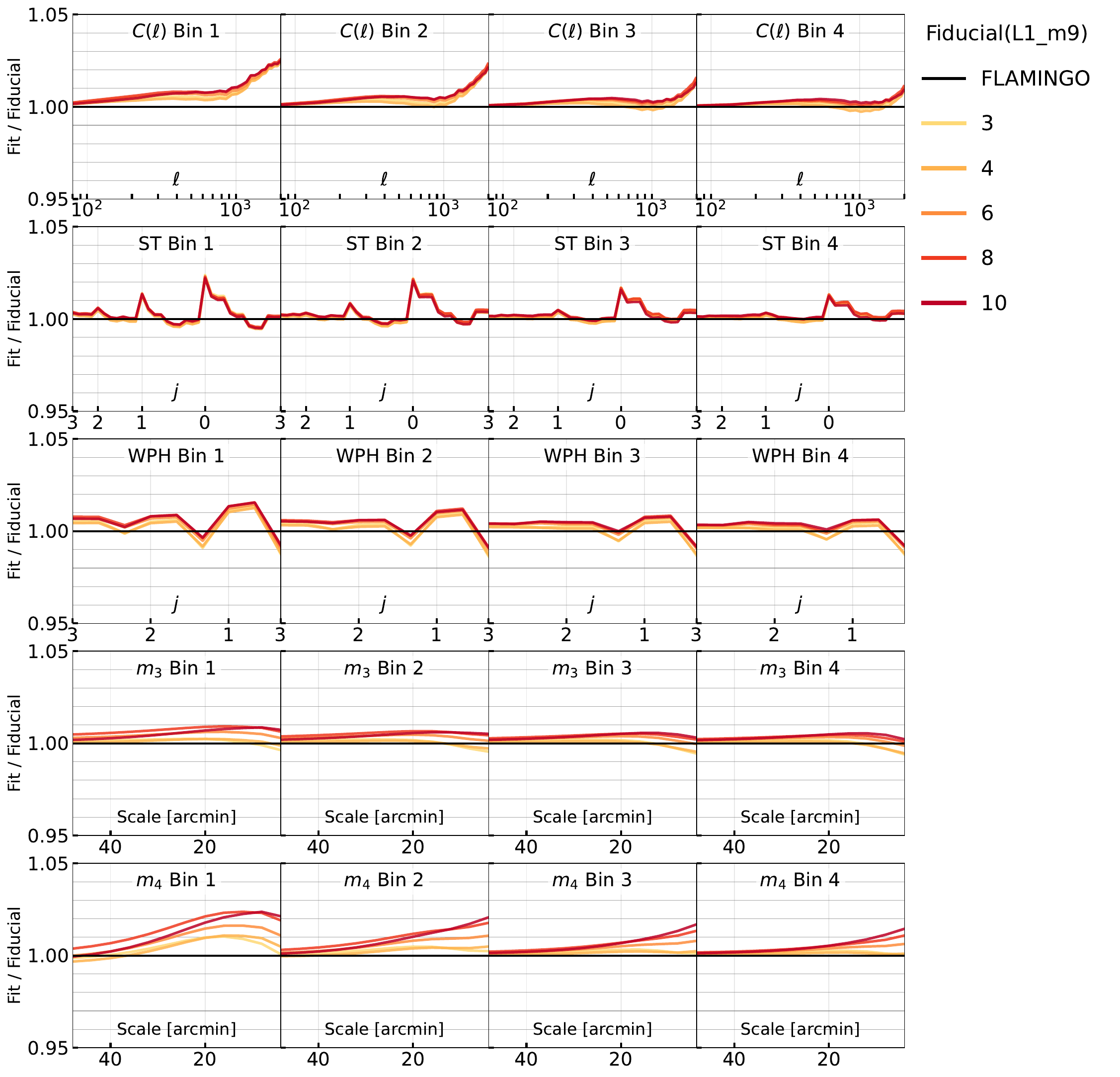}
    \caption{The residuals error of the BCM when fitted to the \flamingo{} fiducial variant. Rows correspond to different statistics, and columns to redshift bins. The y-axis shows the ratio between the statistics of the best-fit baryonified map to that of the fiducial variant. On the x-axis, spatial scales decreases to the right. ST and WPH are ordered by the wavelet size ($j$) as a proxy for their characteristic scale. Colors indicate the number of free parameters in the BCM fit. The shaded region marks $\pm$2\% for visual reference. For all n-parameter models, the BCM flexibly reproduces the baryonic features in the \flamingo{} fiducial variant simulation to percent level. The prediction is robust even with only 3 degrees of freedoms. Differences in accuracy between the models stem from how we estimate the best-fit parameter, as discussed in \Sec\ref{sec:reduced_param}. The residuals for the strongest (fgas$-8\sigma$) and weakest (fgas$+2\sigma$) feedback variants are shown in \Fig\ref{fig:feedback+} and \Fig\ref{fig:feedback-} in \App\ref{appx:residuals}.}
    \label{fig:feedback_fiducial}
\end{figure}

\section{Results}\label{sec:result}
\subsection{The 10-parameter baryon correction model}

We sample the posterior $\mathcal{P}(\bary\mid\stat_{\mathbf{hydro}})$ across the full 10-dimensional baryon parameter space for each of the \flamingo{} feedback variants. Since the marginal posterior distributions are approximately Gaussian, we use the medians -- rather than the maximum a posteriori (MAP) values, which are less robust against sample variance -- as our best-fit parameters $\barybf$. The estimated $\barybf$ values and their 1$\sigma$ uncertainties are listed in \Tab\ref{tab:all_parameter_constraints} in \App\ref{appx:table_bestfit}.

We then investigate how well the best-fit BCM reproduces the statistics in $\kappahydro$. Ideally, we would directly compare the emulator prediction $r(\barybf)$ with \flamingo{}. However, the emulator includes bias from both the neural network and the BCM itself. To isolate the latter, we directly construct $\kappabary(\barybf)$, measure its statistics, and compare them to the \flamingo{} results. The emulator results agree with direct baryonification within expected emulator errors. For example, for \flamingo{}’s fiducial variant, the relative error of the 10-parameter best-fit BCM is shown in \Fig\ref{fig:feedback_fiducial} in dark red. The BCM accurately reproduces ST, WPH, $C(\ell)$, and moment statistics at all scales and redshifts to within 2\%. The residual error is the largest at small scales and low redshifts, where baryonic feedback is strongest.

 We analyze all 8 other \flamingo{} variants in the same way and show the residuals for the strongest (fgas$-8\sigma$) and weakest (fgas$+2\sigma$) \flamingo{} feedback variants in \Fig\ref{fig:feedback+} and \Fig\ref{fig:feedback-} in \App\ref{appx:residuals}. The BCM can reproduce the statistics of different \flamingo{} variants equally well. For the strongest AGN feedback case (fgas$-8\sigma$), the BCM shows elevated errors of 3–4\% in the first redshift bin of $m_4$ for $\theta < 20$ arcminutes,  but otherwise maintains 2\% accuracy. However, as discussed in \Sec\ref{sec:stat_mom}, higher-order moments are intrinsically noisier in both modeling and measurement. The observed $m_4$ discrepancy remains well below the statistical uncertainty of a DES-like survey \citep{gattiDarkEnergySurvey2024} and is only detected at low significance in current DES Y3 data \citep{Anbajagane:2023:CDFs}.

We have verified that a significant fraction of the remaining 2\% residual at small scales can be alleviated by lowering the halo mass threshold to \(10^{12.5} \, M_{\odot}/h\). This test was performed using the best-fit of the baryonification emulator on the Fiducial simulation, and should be considered an approximate check. A formal evaluation would require regenerating the maps, retraining the emulator, and refitting the model using \(10^{12.5} \, M_{\odot}/h\) as the threshold — a process we do not pursue due to its substantial computational cost. Nonetheless, we note that adopting a threshold of \(10^{13} \, M_{\odot}/h\) still yields sub-2\% precision.

Thus far, we have investigated the BCM at the fiducial cosmology. While both $\stat_{\mathbf{hydro}}$ and $\stat_{\mathbf{dmo}}$ depend on cosmology and astrophysics, these dependencies are approximately factorizable \citep{vandaalenEffectsGalaxyFormation2011,elbersFLAMINGOProjectCoupling2025}. As a result, the cosmology dependence largely cancels in the ratio $\stat_{\mathbf{hydro}} / \stat_{\mathbf{dmo}}$, which the BCM models. Using the low-$S_8$ variants (LS8, LS8\_fgas-8$\sigma$) of the \flamingo{} suite, we show in \App\ref{appx:cosmology} that this ratio varies by less than 1\% compared to the fiducial cosmology even for the strong feedback model (fgas-8$\sigma$). Since the BCM (and thus the emulator) predicts the suppression ratio solely from the baryonic parameters, the recovered LS8 BCM parameters are consistent with those from the fiducial cosmology. 
The baryonic suppression has a stronger dependence on the baryon fraction, $\Omega_b / \Omega_m$ \citep[e.g.][]{schneiderQuantifyingBaryonEffects2019} but we leave a detailed analysis of this dependence to future work due to the computational cost of training separate emulators for each cosmology.

To summarize, for $\ell \leq 2000$, the 10-parameter BCM is flexible enough to reproduce the range of baryonic signatures in two-point and HOS. 

\subsection{Baryon correction model with restricted degrees of freedoms}\label{sec:reduced_param}
\input{table_median_v5_3param}
After validating that the 10-parameter BCM accurately reproduces the statistics, we ask whether similar success can be achieved with significantly fewer degrees of freedom. This simplification has three advantages. First, in real survey analyses, using fewer baryon parameters reduce the impact on the constraining power of key cosmological parameters. Second, in simulation-based inference, it lowers the number of simulations required to train the neural posterior estimator, thus reducing computational cost. Third, the 10-dimensional $\bary$ space contains significant degeneracies; fewer parameters improves the interpretability how $\bary$ vary across \flamingo{} feedback variants. 

We consider 4 restricted versions of the BCM, with 8, 6, 4, and 3 parameters. In general, at each step, we fix parameters whose best-fit values are consistent across all feedback variants. The 8-parameter model fixes the redshift scalings $\nu_{M_c}$ and $\nu_{\thetaej}$ to zero, consistent with their posteriors in the 10-parameter fits. The 6-parameter model further fixes $\gamma = 2.6$ and $\log_{10} \theta_{co} = -1$. For the 4-parameter model, we also fix $\log_\delta \eta_\delta = 0.28$ and $\delta = 14$. Finally, for the 3-parameter model, we additionally fix $\mu_\beta = 1.6$. The remaining free parameters in the 3-parameter model are $\log_{10} M_c$, $\log_{10} \eta$, and $\thetaej$. For each n-parameter model and feedback variant, we tabulate $\barybf$ in \Tab\ref{tab:all_parameter_constraints} in \App\ref{appx:table_bestfit}, with fixed parameters shown in red. We also summarize $\barybf$ values for the 3-parameter model in \Tab\ref{tab:3_parameter_constraints}. As with the 10-parameter case, we validate the best-fit BCMs by directly constructing $\kappabary$ and measuring their statistics. Residuals for the fiducial, strong feedback (fgas$-8\sigma$), and weak feedback (fgas$+2\sigma$) variants are shown in \Figs\ref{fig:feedback_fiducial}, \ref{fig:feedback+}, and \ref{fig:feedback-}, respectively.

Although we have dramatically reduced the number of parameters, all n-parameter models faithfully reproduce all statistics across scales, redshifts, and feedback variants. Notably, the 3-parameter model often fits the statistics better than the 10-parameter model. This is because our best estimate, $\barybf$, is defined as the median of the marginal posterior. In high-dimensional spaces, $\barybf$ can deviate from the MAP due to projection over non-Gaussian volumes. As the number of parameters decreases, the posterior becomes more Gaussian-like, and $\barybf$ approaches the MAP, leading to improved fits. In other words, were we to have a less noisy emulator, we could potentially use the MAP as $\barybf$ and obtain even higher-quality fits for the 10-parameter model.

\subsection{Baryon parameters}\label{sec:parameters}
\begin{figure}
    \centering
    \includegraphics[width=0.8\linewidth]{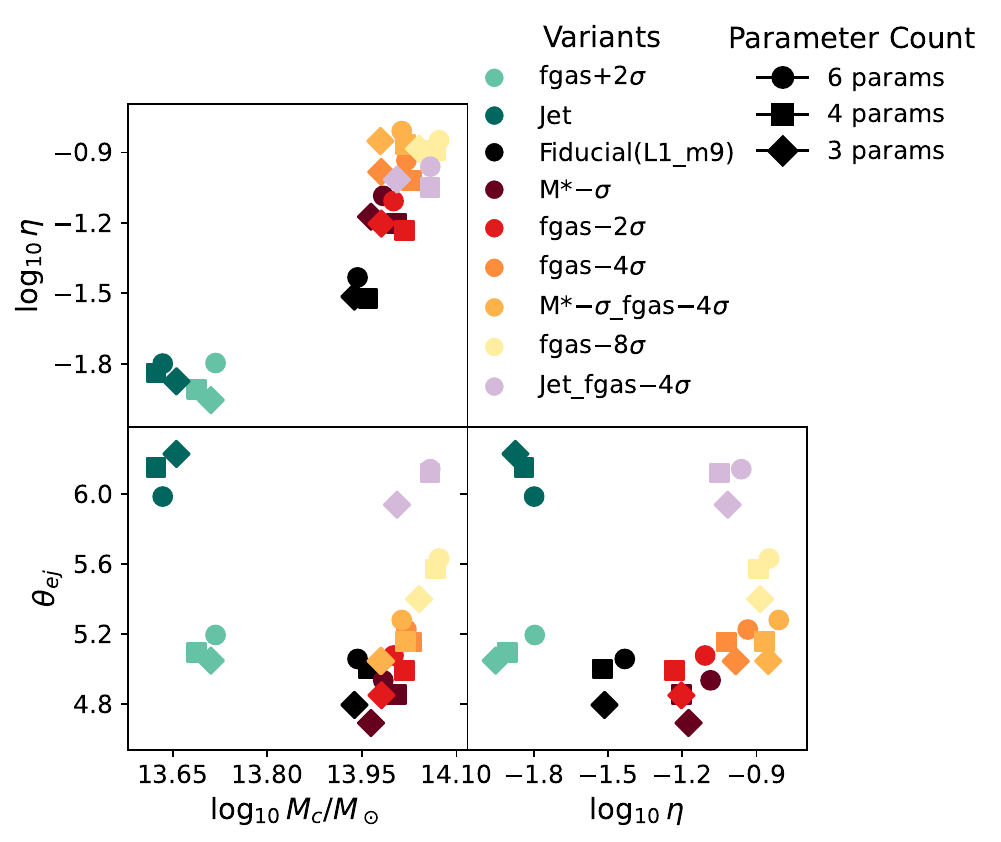}
    \caption{The inferred baryon parameter $\barybf$ from different \flamingo{} feedback variants. Markers represent inference using different degrees of freedom. The color map labels the feedback variants and inherits from \Fig\ref{fig:data}. In the legend, the feedback strength approximately increases from top to bottom, with the strongest (weakest) feedback model suppression being the fgas$-8\sigma$ (fgas$+2\sigma$) model. For each feedback variant, the clustering of the points demonstrate the BCM is robust against removing many nuisance parameter dependencies. }
    \label{fig:parameters}
\end{figure}

Having demonstrated that the BCM can flexibly reproduce the baryonic signatures across a range of \flamingo{} feedback variants, we now examine the best-fit baryon parameters themselves. When comparing with other results in the literature, we emphasize that our modeling choices — such as the specific implementation of the BCM and the set of free parameters —  are in some aspects different from those adopted in other studies, and that our results are entirely based on simulations. In this context, consistency at the parameter level with earlier work is not required for our approach to be effective, as success is defined by how well it reproduces the suppression of the lensing statistics in the FLAMINGO simulations, rather than by agreement in individual parameter values. Nonetheless, the inferred $\barybf$ is broadly consistent with earlier results based on both simulations and observations \citep{schneiderQuantifyingBaryonEffects2019, giriEmulationBaryonicEffects2021, bigwoodWeakLensingCombined2024}. This qualitative agreement reinforces that our model is not overfitting to a narrow region of parameter space and is instead capturing physically meaningful baryonic effects.

Focusing on the 10-parameter model, all variants are consistent with $\nu_{M_c} = 0$ and $\nu_{\thetaej} = 0$, suggesting minimal sensitivity of our measurements to the mass dependence of the gas ejection radius and characteristic slope. This lack of sensitivity is consistent with the forecasted constraints from weak lensing presented in \cite{anbajaganeMaplevelBaryonificationEfficient2024}. Interestingly, all feedback variants prefer a high $\delta$, which corresponds to a sharp drop off in the gas density profile past $r > \theta_{\rm ej}\Rtwohc$ \citep{anbajaganeMaplevelBaryonificationEfficient2024,schneiderQuantifyingBaryonEffects2019,bigwoodWeakLensingCombined2024}. To investigate if different values of $\delta$ affect the best-fit statistics, we rerun the 4- and 3-parameter models, setting $\delta=7$ and $\mu_\beta = 2$ while keeping the other fixed parameters the same as in \Sec\ref{sec:reduced_param}. 
The BCM still achieves a $2\%$ fit across all statistics and feedback variants, although the best-fit $\thetaej$ lowers from $5$ to $3$. This is expected as $\delta$ and $\thetaej$ both control the radial extent of the gas profile and are therefore quite negatively correlated. 

Next, we investigate how much the estimated $\barybf$ varies across the n-parameter models. \Fig\ref{fig:parameters} shows $\barybf$ as a function of feedback variant for the 6-, 4-, and 3-parameter models. The constraints on the remaining free parameters do not vary significantly as we reduce the number of parameters. This demonstrates that the BCM's {weak lensing predictions} are robust against removing many nuisance parameters and the BCM allows us to track how the phenomenological baryon parameters change as a function of \flamingo{} feedback models. The best-fit parameters of the different models show some degree of clustering in the 3D parameter space, indicating that, if sufficient statistical power is achieved by observational probes, it may be possible to distinguish between feedback models using real data. Concerning the physical interpretation of the position of the models in the parameter space, we know that $\log_{10} M_c$, $\log_{10} \eta$, and $\theta_{\rm ej}$ all scale positively with feedback strength. Indeed, we find that the strongest feedback model (fgas$-8\sigma$) lies in the upper range of each of the three parameters, while the weakest feedback model (fgas$+2\sigma$ weak) occupies the lowest values in each dimension. Beyond a qualitative ranking of suppression strength, the interplay between the three parameters and the resulting impact on the lensing statistics is non-trivial. Each parameter modulates the suppression differently and in a scale-dependent way, leading to best-fit values that are scattered across parameter space in a complex pattern.

%% file: table_median_v5_3param.tex
\small
\begin{table}[ht]
\centering
\begin{tabular}{lccc}
\toprule
\textbf{Variant} & $\log_{10} M_{\rm c} / \Msun$ & $\log_{10}\eta$ & $\thetaej$ \\
\midrule 
Fiducial(L1\_m9) & $13.9 \pm 0.19$ & $-1.51 \pm 0.39$ & $4.80 \pm 0.65$ \\
Jet & $13.7 \pm 0.16$ & $-1.87 \pm 0.39$ & $6.23^{+0.89}_{-0.75}$ \\
Jet\_fgas$-4\sigma$ & $14.0 \pm 0.16$ & $-1.02^{+0.21}_{-0.25}$ & $5.94 \pm 0.61$ \\
fgas$+2\sigma$ & $13.7 \pm 0.16$ & $-1.95 \pm 0.38$ & $5.05^{+0.68}_{-0.59}$ \\
fgas$-2\sigma$ & $14.0 \pm 0.22$ & $-1.20^{+0.34}_{-0.35}$ & $4.85 \pm 0.59$ \\
fgas$-4\sigma$ & $14.0 \pm 0.19$ & $-0.983^{+0.23}_{-0.29}$ & $5.05 \pm 0.54$ \\
fgas$-8\sigma$ & $14.0 \pm 0.14$ & $-0.885^{+0.17}_{-0.18}$ & $5.40 \pm 0.52$ \\
M*$-\sigma$ & $14.0 \pm 0.21$ & $-1.17^{+0.33}_{-0.37}$ & $4.69 \pm 0.55$ \\
M*$-\sigma$\_fgas$-4\sigma$ & $14.0 \pm 0.15$ & $-0.852 \pm 0.2$ & $5.05 \pm 0.5$ \\
\bottomrule
\end{tabular}
\caption{The best-fit BCM parameters obtained from the joint fitting of all HOS to different \flamingo{} variants. These parameter combinations yield HOS that agree with \flamingo{} measurements to 2\%. Other BCM parameters are fixed to $\nu_{M_c}=0$, $\nu_{\thetaej}=0$, $\gamma=2.60$, $\log_{10}\theta_{co}=-1.00$, $\log_{10}\eta_{\delta}=-0.28$, $\delta=14.0$, and $\mu_\beta=1.60$. The uncertainty is the $16\%$ and $84\%$ credible interval of our fitting method. See \Tab\ref{tab:all_parameter_constraints} in \App\ref{appx:table_bestfit} for the best-fit values for all n-parameter models.}
\label{tab:3_parameter_constraints}
\end{table}
\normalsize

%% file: 7_conc.tex
\section{Conclusion}\label{sec:conc}
Data from current and upcoming weak lensing surveys are probing cosmological structures at increasingly small spatial scales. To extract both cosmological and astrophysical information in this regime robustly and optimally, we require models that can accurately capture the full range of baryonic feedback effects allowed by theory and observed in state-of-the-art hydrodynamical simulations.

In this work, we investigate the map-level baryonification approach from \citet{anbajaganeMaplevelBaryonificationEfficient2024}, which efficiently and flexibly models baryonic effects in weak lensing convergence maps. By calibrating the baryon parameters against \flamingo{} hydrodynamical simulations, we demonstrate that this baryon correction model (BCM) accurately reproduces baryonic impacts on various convergence field statistics, including the two-point power spectrum, scattering transform coefficients (ST), wavelet phase harmonics (WPH), and third- and fourth-order moments. This agreement holds across scales $\ell < 2000$ and throughout the redshift range of DES-like surveys. We construct an emulator to predict BCM-generated statistics efficiently, then sample baryon parameters by comparing emulator predictions with measurements from different \flamingo{} feedback variants. Testing BCM versions with 10, 8, 6, 4, and 3 free parameters, we find all models can reproduce statistics within $<$ 2\% accuracy with accuracy improving at larger scales and higher redshifts. A minor exception is the fourth moment, where, for models with more than 3 free parameters, the residual error approaches $3\%$ around $10 \arcmin$, partially due to the increased uncertainty in our fitting procedure.

These results have several implications for future weak lensing analyses. First, this shows that the BCM is robust for a wide range of weak lensing statistics and can be embedded in simulation-based inference frameworks to marginalize over baryonic uncertainties or directly constrain baryon parameters. Reduced-parameter versions of the BCM further improve efficiency by lowering the number of simulation needed and sampling costs. Second, our work can be extended to test the robustness of other tracers and probes, such as cross-correlations with large-scale structure observables like galaxy clustering or thermal Sunyaev-Zel’dovich maps. Joint analyses with these additional datasets may offer tighter constraints on baryonic physics.

Although our results are based on the DES Y3 redshift distributions, they are broadly applicable to other current weak lensing surveys due to similar source redshift distributions \citep{Rau:2023:Redshifts, Anbajagane:2025:DECADERedshifts, Wright:2025:LegacyRedshifts}. They are also relevant for upcoming surveys like LSST, Euclid, and Roman, whose higher-redshift distributions may allow our models to extend to even smaller angular scales due to weaker baryonic feedback at early times. The ability to model baryonic physics in this regime opens new opportunities for extracting both cosmological and astrophysical information from small-scale data. Future work could extend this framework to alternative summary statistics, such as voids, peaks, or CNN-based approaches \citep[e.g.,][]{jainStatisticsDarkMatter2000, jeffreyDarkEnergySurvey2024a}. Our results highlight the promise of map-level baryonification—without relying on 3D particle snapshots—as a computationally efficient and physically grounded approach for incorporating baryonic feedback in weak lensing analyses. As upcoming surveys push to smaller scales, such methods will be essential for maximizing constraining power while remaining robust to astrophysical systematics.

%% file: main_appendix.tex
\appendix
\section{Scale cut}
\label{appx:scale}
\begin{figure}
    \centering
    \includegraphics[width=1.0\linewidth]{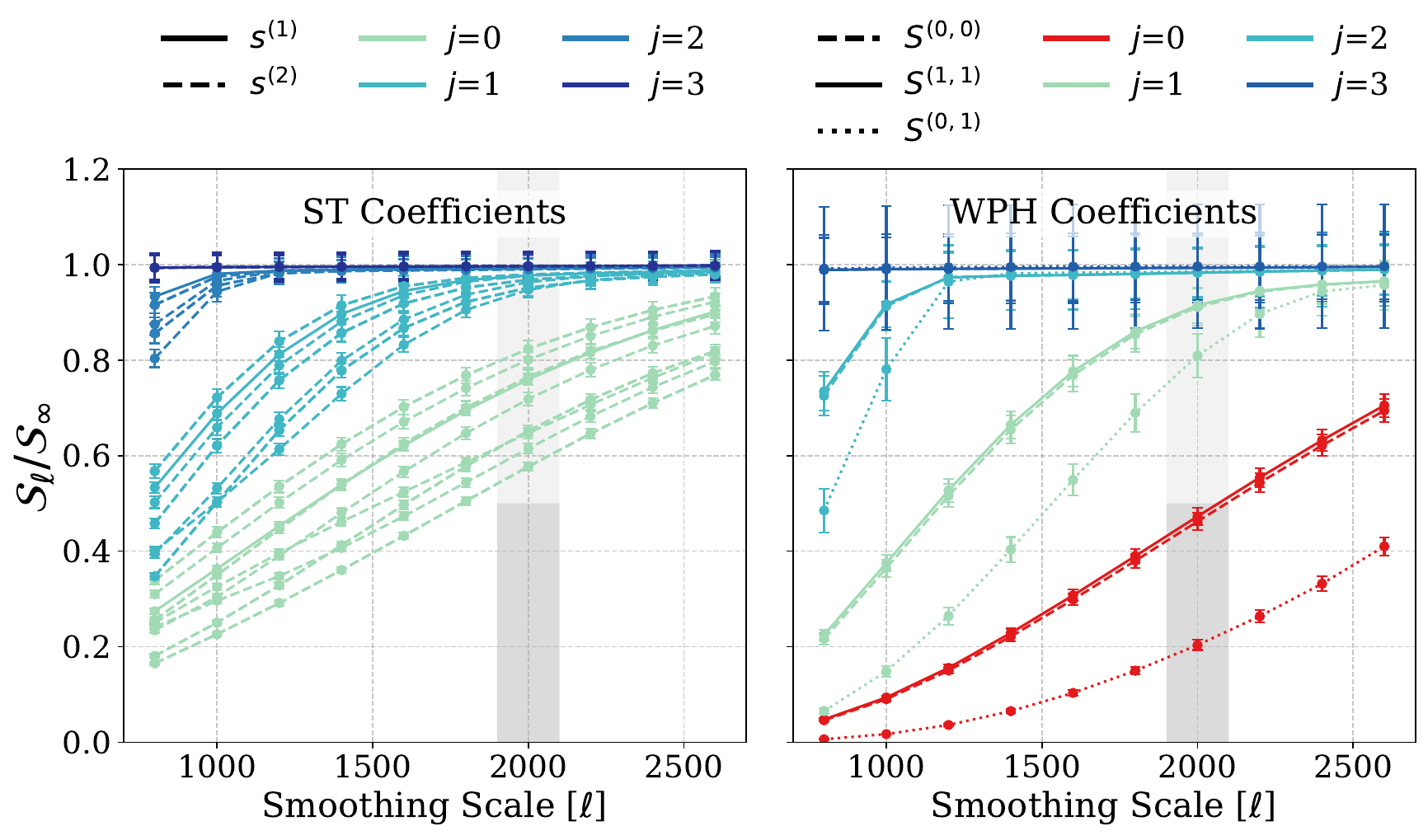}
    \caption{The effect of smoothing scale on the two kinds of wavelet coefficients, ST and WPH. Given a statistics $\stat$, we define $\stat_\ell$ as the statistics measured on $\kappa$ maps that are low-pass filtered (with a step function) at different $\ell$. The statistics measured on the smoothed map is called $\stat_\infty$. We test this idea on $\kappadmo$, and plot the relative suppression $\stat_\ell / \stat_\infty$ as a function of smoothing scale $\ell$. The line and the error bar represent the mean and the standard error of $\stat$ measured on different patches across the full sky. We further color code the lines using the characteristic scale of the wavelet (higher $j$ equals larger wavelets). If $\stat_\ell / \stat_\infty \ll 1$, then $\stat$ contains information beyond $\ell$. The $j=0$ WPH coefficients (shown in red) are suppressed more than 50\% at $\ell=2000$ (shaded region), and hence removed from subsequent analyses.}
    \label{fig:scale_cut}
\end{figure}
We empirically determine the characteristic scale of each ST and WPH statistic and enforce a consistent scale cut with respect to the power spectrum. For each statistic $\stat$, we smooth the $\nside=1024$ $\kappadmo$ maps with low-pass filters at different $\ellcut$ values and measure the statistics ($\stat_\ell$), normalized by the unsmoothed map’s statistics ($\stat_\infty$). The results for ST and WPH are shown in the left and right panels of \Fig\ref{fig:scale_cut}, respectively. For a given statistic, if $\stat_\ell / \stat_\infty \ll 1$, then it contains information beyond $\ell$, and vice versa. We exclude statistics where $\stat_\ell / \stat_\infty < 0.5$ at $\ellcut = 2000$. All 34 ST coefficients per redshift bin remain within our analysis scales. For WPH, the three $j=0$ coefficients (shown in red) fail the scale cut criterion and are removed from the analysis.

\section{Statistics as a function of baryon parameters}
\label{appx:sensitivity}
\begin{figure}
    \centering
    \includegraphics[width=1\linewidth]{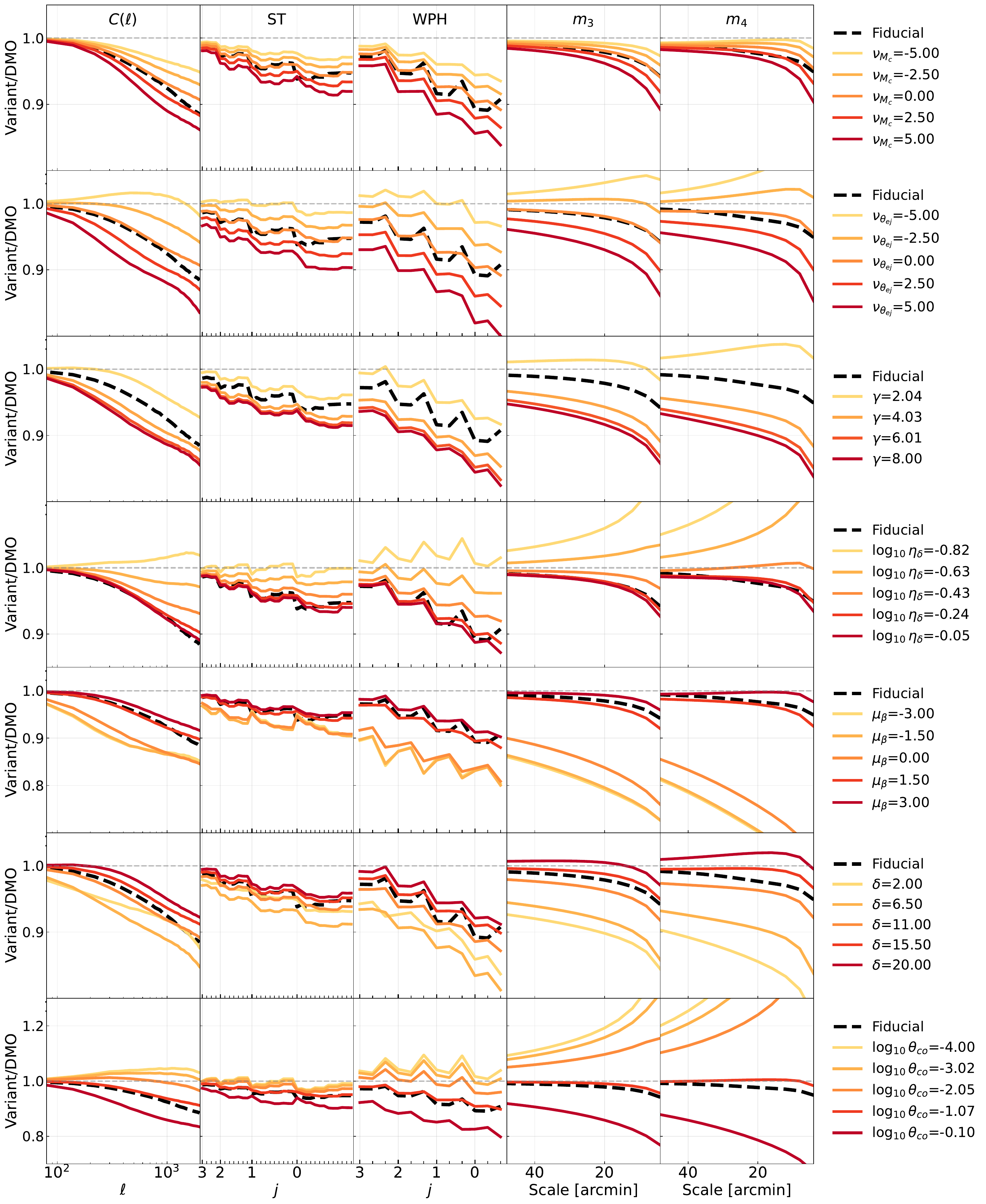}
    \caption{Same as \Fig\ref{fig:variation_3param}, but for the remaining BCM parameters not shown in the main text. Colored curves show the summary statistics of baryonified convergence maps with one parameter varied at a time, while black dashed curves show the corresponding statistics from the \flamingo{} simulations. }
    \label{fig:variation_7param}
\end{figure}
Here we show how the summary statistics vary with the remaining BCM parameters listed in \Tab\ref{tab:params_descrp}, excluding $M_c$, $\theta_{\rm ej}$, and $\eta$, which are highlighted in the main text (\Fig\ref{fig:variation_3param}). Each parameter is varied individually across its prior range, and the resulting baryonified statistics are compared to those from the corresponding DMO simulation. These results show the flexibility of the BCM in reproducing a wide range of baryonic effects around the hydrodynamical predictions.

\section{Emulator}
\label{appx:emulator}
We build a neural network-based emulator to map the baryon parameters to the tomographic ST, WPH, $C(\ell)$ and the $m_3$ and $m_4$ statistics. Each type of statistic is modeled independently. We first standardize the input parameters and then use three dense hidden layers following the architectures: (512, 256, 256) for ST, (512, 256, 128) for WPH, (512, 256, 128) for $C(\ell)$ statistics, and (512, 256, 128) for the concatenated $m_3$ and $m_4$ data vector. Each hidden layer is followed by a leaky ReLU activation function ($\alpha=0.1$) \citep{xuEmpiricalEvaluationRectified2015} and a 3\% dropout layer to prevent over-fitting \citep{JMLR:v15:srivastava14a}. For robustness, we train 10 independent cross-validated models for each type of statistics, resulting in a total ensemble of $40$ neural networks. We use $95\%$ of the data for the 10-fold cross-validation and use the rest as the test set for the performance evaluation as shown in \Fig\ref{fig:emulator}. During training, we use the mean absolute error (MAE) as the loss function and the Adam optimizer with weight decay for $l_2$ regularization \citep{kingmaAdamMethodStochastic2017}. The emulator implements a dual early stopping strategy to prevent over-fitting: (1) a global early stopping mechanism with a patience of 80 iterations that monitors validation loss to terminate training when no improvement is observed, and (2) a learning rate scheduler that starts with a learning rate of $10^{-3}$ and reduces it by a factor of 0.5 after 5 epochs without improvement, with a minimum threshold of $10^{-6}$. Since each statistic is predicted 10 times by the cross-validated models, our emulator not only provides a point estimate (given by the ensemble mean) but also uncertainty estimates (given by the ensemble variance). The entire implementation supports automatic differentiation through \jax, enabling gradient-based optimization and sampling, as well as sensitivity analysis of the statistics with respect to the baryon parameters \citep{jax2018github}.

We used the baryonification model to generate $6695$ $\kappabary(\bary)$ maps and measured their statistics $\statbary$, where the 10-dimensional $\bary$ is sampled using a Sobol sequence. Since some $\statbary$ show an amplification rather than a suppression as observed in all \flamingo{} measurements, we remove $\statbary$ where $C(\ell)$ is on average amplified by $10\%$ across all linear bins. The final training set includes $5468$ $\statbary$. Since the training set is small, the emulator has high outlier rate near parameter boundaries. Therefore, we truncate parameter domain boundaries where the absolute error averaged across all statistics is above $1\%$. The most significant truncation occurs for $\gamma < 0.9$. We later verify these restrictions do not impact the posteriors of the MCMC chains.

\input{table_median_v5}

\clearpage
\section{Baryonification error for strong and weak feedback variants}
\label{appx:residuals}
\begin{figure}[ht]
    \centering
    \includegraphics[width=1\textwidth]{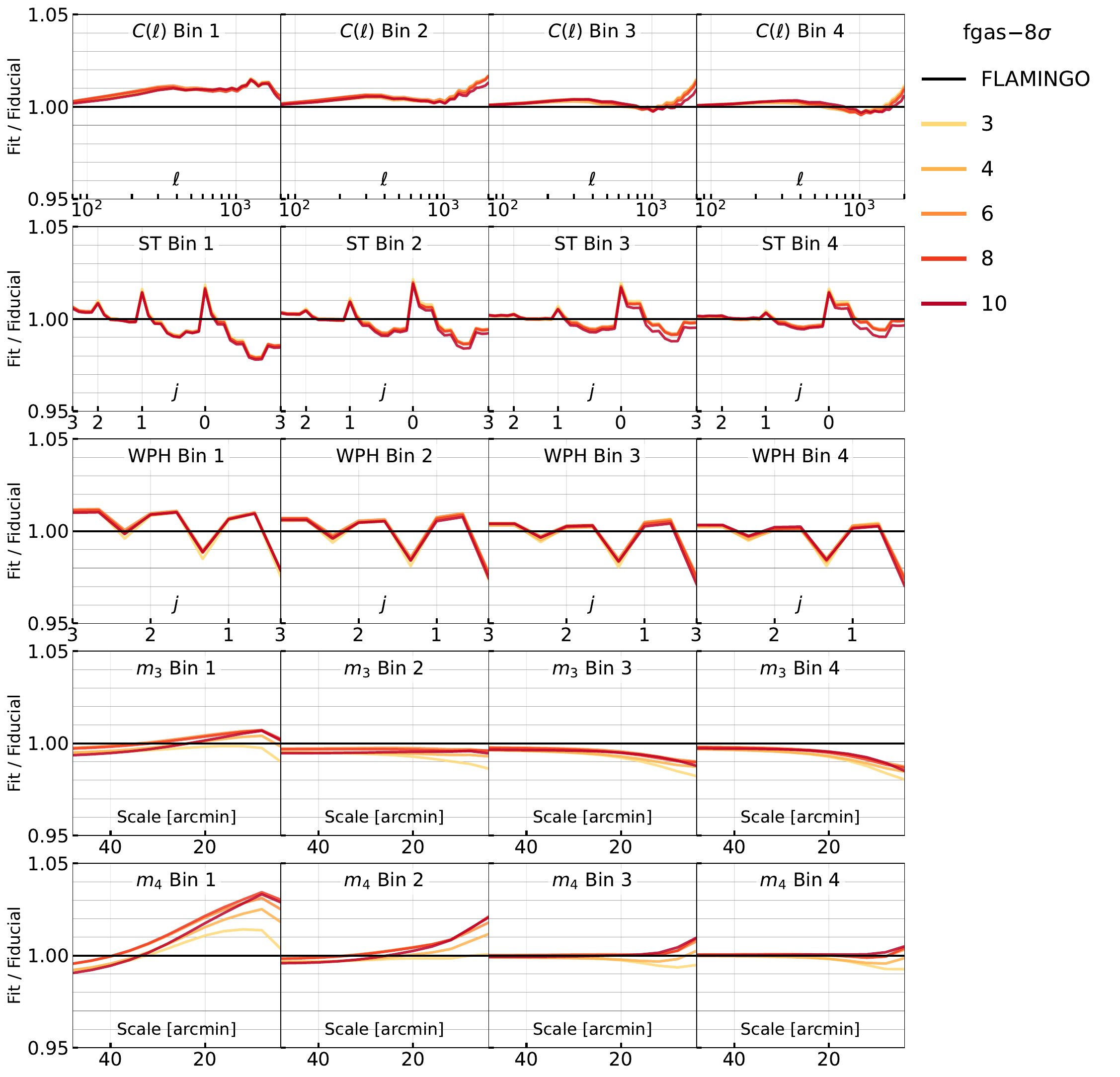}
    \caption{Similar to \Fig\ref{fig:feedback_fiducial} but with BCM fitted to the strongest feedback variant in the \flamingo{} suite. The BCM error is still at the percent level. The elevated residual for low redshift $m_4$ is potentially related to both the baryon modelling error and/or $m_4$'s sensitivity to noise.}
    \label{fig:feedback+}
\end{figure}

\begin{figure}
    \centering
    \includegraphics[width=1\textwidth]{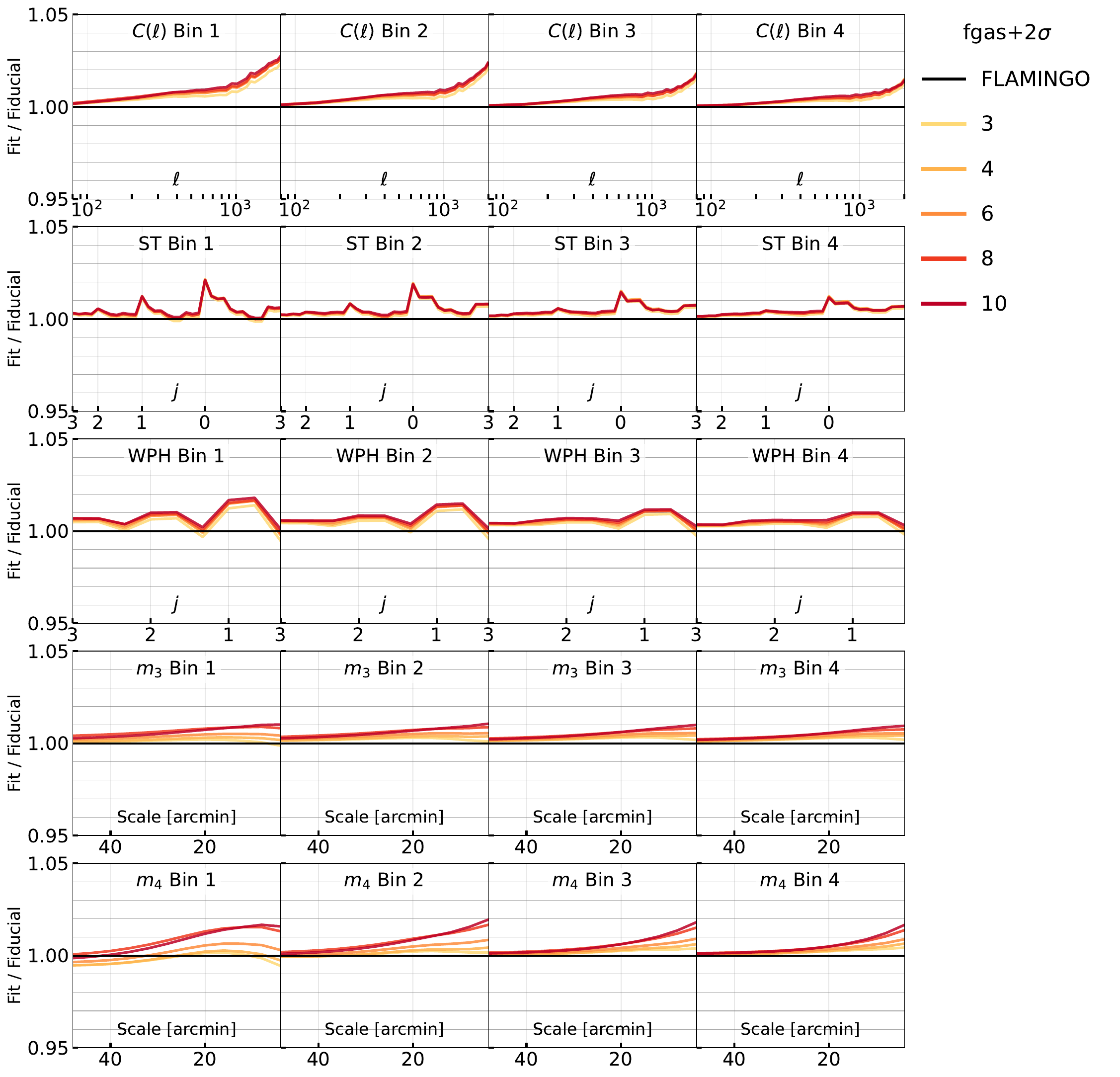}
    \caption{Similar to \Fig\ref{fig:feedback_fiducial} but with BCM fitted to the weakest feedback variant in the \flamingo{} suite. The BCM error is still at the percent level.}
    \label{fig:feedback-}
\end{figure}

Following the discussion in \Sec\ref{sec:result}, we show here the residual errors for the best-fit BCM models applied to two extreme feedback variants from the \flamingo{} suite: the strongest (fgas$-8\sigma$) and the weakest (fgas$+2\sigma$). These figures are analogous to \Fig\ref{fig:feedback_fiducial}, which showed results for the fiducial case. For the strong feedback case (fgas$-8\sigma$), the BCM reproduces most statistics to within 2\%, with a mild 3–4\% excess in $m_4$ at low redshift and small angular scales (\Fig\ref{fig:feedback+}). This reflects both modeling limitations and the intrinsic noise in higher-order moments. For the weak feedback case (fgas$+2\sigma$), residuals remain within 2\% across all statistics and scales (\Fig\ref{fig:feedback-}).

\section{Cosmology dependence of baryonic suppression}
\begin{figure}
    \centering
    \includegraphics[width=0.95\linewidth]{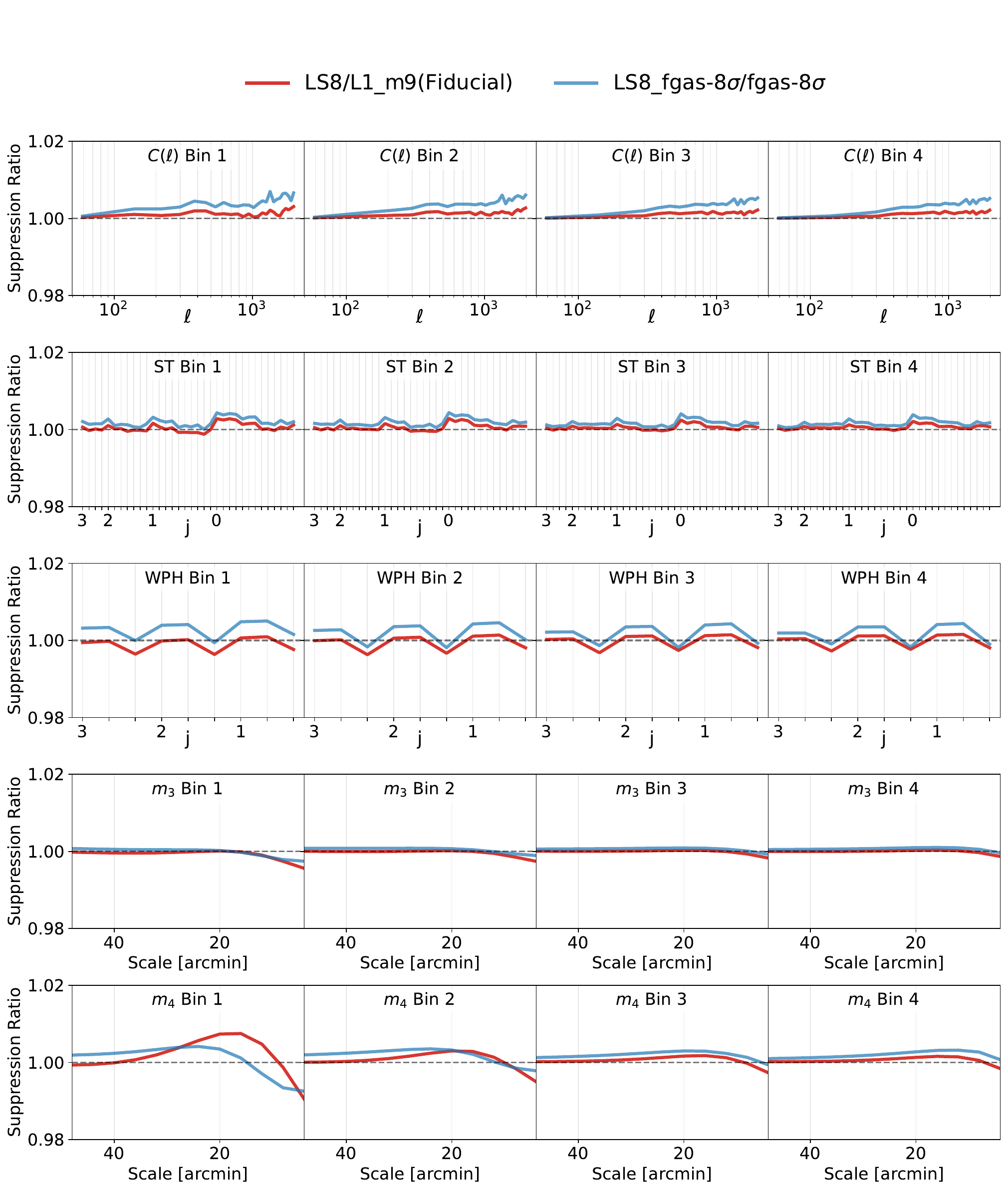}
    \caption{The sensitivity of the baryonic effect to cosmology for the first redshift bin. The y-axis shows the fractional change between the baryonic suppression of summary statistics at the LS8 cosmology compared to the suppression at the fiducial cosmology, for two different subgrid models: the fiducial models (LS8 and L1\_m9) and the strong AGN feedback variants (LS8\_fgas-8$\sigma$ and fgas-8$\sigma$). Across both feedback variants, the amount of baryonic suppression changes only at the sub-percent level, indicating weak dependence on $\sigma_8$.}
    \label{fig:cosmology}
\end{figure}

To test how baryonic suppression of weak lensing summary statistics varies with cosmology, we compute the statistics for both the fiducial and LS8 cosmologies, and for each with the fiducial and strong-AGN (fgas$-8\sigma$) feedback variants. \Fig\ref{fig:cosmology} shows the relative difference in suppression between LS8 and fiducial cosmologies in the first redshift bin, which is sub-percent for all statistics and scales.

\label{appx:cosmology}

%% file: table_median_v5.tex
\clearpage
\newgeometry{left=1in,right=1in,top=1in,bottom=1in}
\begin{landscape}
\section{Best-fit baryon parameters}
\label{appx:table_bestfit}
\vspace{0.5em}

\small

\begin{longtable}{lcccccccccc}
\caption{The best-fit BCM parameters obtained from the joint fitting of all HOS to different \flamingo{} variants. These parameter combinations yield HOS that agree with \flamingo{} measurements to 2\%. The integers following the feedback variant names are the number of free parameters. The fixed parameters are highlighted in red. The uncertainty is the $16\%$ and $84\%$ credible interval. } \\
\label{tab:all_parameter_constraints} \\
\hline
Variant & $\nu_{M_c}$ & $\nu_{\thetaej}$ & $\gamma$ & $\log_{10}\theta_{co}$ & $\log_{10}\eta_{\delta}$ & $\delta$ & $\mu_\beta$ & $\log_{10} M_{\rm c} / \Msun$ & $\log_{10}\eta$ & $\thetaej$ \\ \hline
\endhead

\hline
\multicolumn{11}{r}{Continued on next page} \\
\endfoot

\hline
\endlastfoot

\hline 
Fiducial(L1\_m9)\scriptsize(10) & $1.48^{+2.0}_{-2.4}$ & $-0.727^{+1.1}_{-0.99}$ & $2.61^{+1.7}_{-0.81}$ & $-1.06 \pm 0.19$ & $-0.247^{+0.12}_{-0.16}$ & $13.7 \pm 3.5$ & $1.81 \pm 0.53$ & $13.9 \pm 0.25$ & $-1.57 \pm 0.43$ & $5.36^{+2.4}_{-1.7}$ \\
Fiducial(L1\_m9)\scriptsize(8) & \cellcolor{red!20}$0$ & \cellcolor{red!20}$0$ & $2.64^{+1.8}_{-0.85}$ & $-1.04 \pm 0.18$ & $-0.255^{+0.13}_{-0.18}$ & $14.4^{+3.1}_{-3.7}$ & $1.77 \pm 0.56$ & $13.9 \pm 0.18$ & $-1.49^{+0.48}_{-0.44}$ & $4.85^{+1.8}_{-1.3}$ \\
Fiducial(L1\_m9)\scriptsize(6) & \cellcolor{red!20}$0$ & \cellcolor{red!20}$0$ & \cellcolor{red!20}$2.60$ & \cellcolor{red!20}$-1.00$ & $-0.282^{+0.13}_{-0.17}$ & $15.1^{+2.8}_{-3.9}$ & $1.87^{+0.56}_{-0.48}$ & $13.9 \pm 0.18$ & $-1.43 \pm 0.46$ & $5.06^{+0.91}_{-1.1}$ \\
Fiducial(L1\_m9)\scriptsize(4) & \cellcolor{red!20}$0$ & \cellcolor{red!20}$0$ & \cellcolor{red!20}$2.60$ & \cellcolor{red!20}$-1.00$ & \cellcolor{red!20}$-0.28$ & \cellcolor{red!20}$14.0$ & $1.92^{+0.5}_{-0.42}$ & $14.0 \pm 0.16$ & $-1.52^{+0.37}_{-0.41}$ & $5.00 \pm 0.7$ \\
Fiducial(L1\_m9)\scriptsize(3) & \cellcolor{red!20}$0$ & \cellcolor{red!20}$0$ & \cellcolor{red!20}$2.60$ & \cellcolor{red!20}$-1.00$ & \cellcolor{red!20}$-0.28$ & \cellcolor{red!20}$14.0$ & \cellcolor{red!20}$1.60$ & $13.9 \pm 0.19$ & $-1.51 \pm 0.39$ & $4.80 \pm 0.65$ \\
Jet\scriptsize(10) & $0.697^{+2.4}_{-2.7}$ & $-0.227^{+1.1}_{-1.1}$ & $2.86^{+2.0}_{-0.98}$ & $-0.959^{+0.22}_{-0.21}$ & $-0.305 \pm 0.15$ & $13.5^{+3.8}_{-3.9}$ & $1.50^{+0.72}_{-0.55}$ & $13.6 \pm 0.32$ & $-1.89 \pm 0.41$ & $5.25^{+2.6}_{-1.7}$ \\
Jet\scriptsize(8) & \cellcolor{red!20}$0$ & \cellcolor{red!20}$0$ & $2.92^{+2.0}_{-1.0}$ & $-0.956 \pm 0.2$ & $-0.313 \pm 0.15$ & $14.0^{+3.6}_{-4.2}$ & $1.45^{+0.72}_{-0.58}$ & $13.7 \pm 0.28$ & $-1.86 \pm 0.44$ & $5.01^{+2.1}_{-1.5}$ \\
Jet\scriptsize(6) & \cellcolor{red!20}$0$ & \cellcolor{red!20}$0$ & \cellcolor{red!20}$2.60$ & \cellcolor{red!20}$-1.00$ & $-0.282^{+0.13}_{-0.13}$ & $14.7^{+3.1}_{-4.1}$ & $1.45^{+0.76}_{-0.54}$ & $13.6 \pm 0.27$ & $-1.80^{+0.38}_{-0.41}$ & $5.98 \pm 1.4$ \\
Jet\scriptsize(4) & \cellcolor{red!20}$0$ & \cellcolor{red!20}$0$ & \cellcolor{red!20}$2.60$ & \cellcolor{red!20}$-1.00$ & \cellcolor{red!20}$-0.28$ & \cellcolor{red!20}$14.0$ & $1.49^{+0.74}_{-0.5}$ & $13.6 \pm 0.26$ & $-1.84^{+0.38}_{-0.42}$ & $6.15 \pm 0.98$ \\
Jet\scriptsize(3) & \cellcolor{red!20}$0$ & \cellcolor{red!20}$0$ & \cellcolor{red!20}$2.60$ & \cellcolor{red!20}$-1.00$ & \cellcolor{red!20}$-0.28$ & \cellcolor{red!20}$14.0$ & \cellcolor{red!20}$1.60$ & $13.7 \pm 0.16$ & $-1.87 \pm 0.39$ & $6.23^{+0.89}_{-0.75}$ \\
Jet\_fgas$-4\sigma$\scriptsize(10) & $1.79^{+1.9}_{-2.3}$ & $-0.785^{+1.2}_{-1.0}$ & $2.70^{+1.6}_{-0.77}$ & $-1.08 \pm 0.15$ & $-0.249^{+0.13}_{-0.23}$ & $13.5^{+3.4}_{-3.2}$ & $1.93 \pm 0.39$ & $14.0 \pm 0.2$ & $-1.13^{+0.34}_{-0.37}$ & $6.84^{+2.8}_{-2.0}$ \\
Jet\_fgas$-4\sigma$\scriptsize(8) & \cellcolor{red!20}$0$ & \cellcolor{red!20}$0$ & $2.86^{+1.6}_{-0.84}$ & $-1.07 \pm 0.13$ & $-0.295^{+0.16}_{-0.34}$ & $14.3^{+3.2}_{-3.5}$ & $1.92 \pm 0.37$ & $14.1 \pm 0.16$ & $-0.99^{+0.32}_{-0.35}$ & $5.78^{+1.9}_{-1.3}$ \\
Jet\_fgas$-4\sigma$\scriptsize(6) & \cellcolor{red!20}$0$ & \cellcolor{red!20}$0$ & \cellcolor{red!20}$2.60$ & \cellcolor{red!20}$-1.00$ & $-0.328^{+0.17}_{-0.27}$ & $14.9^{+2.8}_{-3.5}$ & $1.94^{+0.36}_{-0.32}$ & $14.1 \pm 0.14$ & $-0.961^{+0.24}_{-0.31}$ & $6.14 \pm 0.98$ \\
Jet\_fgas$-4\sigma$\scriptsize(4) & \cellcolor{red!20}$0$ & \cellcolor{red!20}$0$ & \cellcolor{red!20}$2.60$ & \cellcolor{red!20}$-1.00$ & \cellcolor{red!20}$-0.28$ & \cellcolor{red!20}$14.0$ & $1.96^{+0.35}_{-0.31}$ & $14.1 \pm 0.13$ & $-1.05^{+0.21}_{-0.26}$ & $6.12^{+0.69}_{-0.62}$ \\
Jet\_fgas$-4\sigma$\scriptsize(3) & \cellcolor{red!20}$0$ & \cellcolor{red!20}$0$ & \cellcolor{red!20}$2.60$ & \cellcolor{red!20}$-1.00$ & \cellcolor{red!20}$-0.28$ & \cellcolor{red!20}$14.0$ & \cellcolor{red!20}$1.60$ & $14.0 \pm 0.16$ & $-1.02^{+0.21}_{-0.25}$ & $5.94 \pm 0.61$ \\
fgas$+2\sigma$\scriptsize(10) & $0.761^{+2.4}_{-2.6}$ & $-0.328^{+1.1}_{-1.1}$ & $2.77^{+2.0}_{-0.9}$ & $-1.07 \pm 0.22$ & $-0.291 \pm 0.15$ & $13.7^{+3.6}_{-4.0}$ & $1.58^{+0.71}_{-0.59}$ & $13.7 \pm 0.29$ & $-1.88 \pm 0.43$ & $5.08^{+2.4}_{-1.7}$ \\
fgas$+2\sigma$\scriptsize(8) & \cellcolor{red!20}$0$ & \cellcolor{red!20}$0$ & $2.75^{+1.9}_{-0.93}$ & $-1.05^{+0.2}_{-0.19}$ & $-0.302 \pm 0.15$ & $14.1^{+3.4}_{-3.9}$ & $1.57^{+0.72}_{-0.59}$ & $13.7 \pm 0.24$ & $-1.85 \pm 0.43$ & $4.82^{+2.1}_{-1.4}$ \\
fgas$+2\sigma$\scriptsize(6) & \cellcolor{red!20}$0$ & \cellcolor{red!20}$0$ & \cellcolor{red!20}$2.60$ & \cellcolor{red!20}$-1.00$ & $-0.317^{+0.13}_{-0.13}$ & $15.0^{+3.2}_{-3.9}$ & $1.66^{+0.75}_{-0.64}$ & $13.7 \pm 0.25$ & $-1.80 \pm 0.43$ & $5.19^{+1.1}_{-1.1}$ \\
fgas$+2\sigma$\scriptsize(4) & \cellcolor{red!20}$0$ & \cellcolor{red!20}$0$ & \cellcolor{red!20}$2.60$ & \cellcolor{red!20}$-1.00$ & \cellcolor{red!20}$-0.28$ & \cellcolor{red!20}$14.0$ & $1.66^{+0.79}_{-0.61}$ & $13.7 \pm 0.25$ & $-1.91 \pm 0.4$ & $5.10 \pm 0.73$ \\
fgas$+2\sigma$\scriptsize(3) & \cellcolor{red!20}$0$ & \cellcolor{red!20}$0$ & \cellcolor{red!20}$2.60$ & \cellcolor{red!20}$-1.00$ & \cellcolor{red!20}$-0.28$ & \cellcolor{red!20}$14.0$ & \cellcolor{red!20}$1.60$ & $13.7 \pm 0.16$ & $-1.95 \pm 0.38$ & $5.05^{+0.68}_{-0.59}$ \\
fgas$-2\sigma$\scriptsize(10) & $1.69^{+2.0}_{-2.2}$ & $-0.802^{+1.0}_{-0.95}$ & $2.59^{+1.5}_{-0.73}$ & $-1.06 \pm 0.16$ & $-0.237^{+0.12}_{-0.19}$ & $13.8^{+3.5}_{-3.3}$ & $1.84 \pm 0.45$ & $13.9 \pm 0.22$ & $-1.31^{+0.43}_{-0.39}$ & $5.53^{+2.3}_{-1.6}$ \\
fgas$-2\sigma$\scriptsize(8) & \cellcolor{red!20}$0$ & \cellcolor{red!20}$0$ & $2.73^{+1.4}_{-0.88}$ & $-1.03 \pm 0.15$ & $-0.267^{+0.14}_{-0.29}$ & $14.5^{+3.2}_{-3.5}$ & $1.83^{+0.45}_{-0.43}$ & $14.0 \pm 0.18$ & $-1.17^{+0.41}_{-0.43}$ & $4.86^{+1.8}_{-1.2}$ \\
fgas$-2\sigma$\scriptsize(6) & \cellcolor{red!20}$0$ & \cellcolor{red!20}$0$ & \cellcolor{red!20}$2.60$ & \cellcolor{red!20}$-1.00$ & $-0.295^{+0.15}_{-0.23}$ & $15.2^{+2.8}_{-3.7}$ & $1.86^{+0.45}_{-0.38}$ & $14.0 \pm 0.17$ & $-1.11^{+0.33}_{-0.42}$ & $5.08^{+0.83}_{-0.93}$ \\
fgas$-2\sigma$\scriptsize(4) & \cellcolor{red!20}$0$ & \cellcolor{red!20}$0$ & \cellcolor{red!20}$2.60$ & \cellcolor{red!20}$-1.00$ & \cellcolor{red!20}$-0.28$ & \cellcolor{red!20}$14.0$ & $1.88^{+0.4}_{-0.36}$ & $14.0 \pm 0.17$ & $-1.23^{+0.32}_{-0.34}$ & $4.99 \pm 0.63$ \\
fgas$-2\sigma$\scriptsize(3) & \cellcolor{red!20}$0$ & \cellcolor{red!20}$0$ & \cellcolor{red!20}$2.60$ & \cellcolor{red!20}$-1.00$ & \cellcolor{red!20}$-0.28$ & \cellcolor{red!20}$14.0$ & \cellcolor{red!20}$1.60$ & $14.0 \pm 0.22$ & $-1.20^{+0.34}_{-0.35}$ & $4.85 \pm 0.59$ \\
fgas$-4\sigma$\scriptsize(10) & $1.68^{+2.0}_{-2.1}$ & $-0.822^{+1.0}_{-0.95}$ & $2.64^{+1.4}_{-0.73}$ & $-1.04 \pm 0.15$ & $-0.24^{+0.13}_{-0.24}$ & $13.8^{+3.3}_{-3.1}$ & $1.89 \pm 0.41$ & $14.0 \pm 0.2$ & $-1.12^{+0.35}_{-0.37}$ & $5.64^{+2.1}_{-1.5}$ \\
fgas$-4\sigma$\scriptsize(8) & \cellcolor{red!20}$0$ & \cellcolor{red!20}$0$ & $2.81^{+1.4}_{-0.83}$ & $-1.02 \pm 0.13$ & $-0.293^{+0.17}_{-0.34}$ & $14.4^{+3.1}_{-3.4}$ & $1.86^{+0.4}_{-0.37}$ & $14.0 \pm 0.16$ & $-0.964^{+0.31}_{-0.36}$ & $4.86^{+1.5}_{-1.0}$ \\
fgas$-4\sigma$\scriptsize(6) & \cellcolor{red!20}$0$ & \cellcolor{red!20}$0$ & \cellcolor{red!20}$2.60$ & \cellcolor{red!20}$-1.00$ & $-0.309^{+0.16}_{-0.25}$ & $15.1^{+2.8}_{-3.6}$ & $1.85^{+0.39}_{-0.34}$ & $14.0 \pm 0.15$ & $-0.935^{+0.25}_{-0.32}$ & $5.23 \pm 0.85$ \\
fgas$-4\sigma$\scriptsize(4) & \cellcolor{red!20}$0$ & \cellcolor{red!20}$0$ & \cellcolor{red!20}$2.60$ & \cellcolor{red!20}$-1.00$ & \cellcolor{red!20}$-0.28$ & \cellcolor{red!20}$14.0$ & $1.88^{+0.37}_{-0.31}$ & $14.0 \pm 0.15$ & $-1.02^{+0.22}_{-0.28}$ & $5.16 \pm 0.58$ \\
fgas$-4\sigma$\scriptsize(3) & \cellcolor{red!20}$0$ & \cellcolor{red!20}$0$ & \cellcolor{red!20}$2.60$ & \cellcolor{red!20}$-1.00$ & \cellcolor{red!20}$-0.28$ & \cellcolor{red!20}$14.0$ & \cellcolor{red!20}$1.60$ & $14.0 \pm 0.19$ & $-0.983^{+0.23}_{-0.29}$ & $5.05 \pm 0.54$ \\
fgas$-8\sigma$\scriptsize(10) & $1.51^{+2.0}_{-2.1}$ & $-0.753^{+1.1}_{-0.93}$ & $2.74^{+1.4}_{-0.76}$ & $-1.04 \pm 0.13$ & $-0.275^{+0.16}_{-0.32}$ & $14.0 \pm 3.2$ & $1.92 \pm 0.38$ & $14.0 \pm 0.19$ & $-0.93^{+0.28}_{-0.3}$ & $5.94^{+2.4}_{-1.6}$ \\
fgas$-8\sigma$\scriptsize(8) & \cellcolor{red!20}$0$ & \cellcolor{red!20}$0$ & $2.96^{+1.4}_{-0.85}$ & $-1.03 \pm 0.12$ & $-0.323^{+0.18}_{-0.36}$ & $14.6^{+3.0}_{-3.4}$ & $1.93 \pm 0.36$ & $14.1 \pm 0.14$ & $-0.857^{+0.26}_{-0.24}$ & $5.12^{+1.5}_{-1.0}$ \\
fgas$-8\sigma$\scriptsize(6) & \cellcolor{red!20}$0$ & \cellcolor{red!20}$0$ & \cellcolor{red!20}$2.60$ & \cellcolor{red!20}$-1.00$ & $-0.322^{+0.17}_{-0.25}$ & $15.0^{+2.7}_{-3.4}$ & $1.91^{+0.35}_{-0.33}$ & $14.1 \pm 0.12$ & $-0.848^{+0.21}_{-0.2}$ & $5.63 \pm 0.83$ \\
fgas$-8\sigma$\scriptsize(4) & \cellcolor{red!20}$0$ & \cellcolor{red!20}$0$ & \cellcolor{red!20}$2.60$ & \cellcolor{red!20}$-1.00$ & \cellcolor{red!20}$-0.28$ & \cellcolor{red!20}$14.0$ & $1.92 \pm 0.35$ & $14.1 \pm 0.12$ & $-0.893^{+0.16}_{-0.19}$ & $5.57 \pm 0.56$ \\
fgas$-8\sigma$\scriptsize(3) & \cellcolor{red!20}$0$ & \cellcolor{red!20}$0$ & \cellcolor{red!20}$2.60$ & \cellcolor{red!20}$-1.00$ & \cellcolor{red!20}$-0.28$ & \cellcolor{red!20}$14.0$ & \cellcolor{red!20}$1.60$ & $14.0 \pm 0.14$ & $-0.885^{+0.17}_{-0.18}$ & $5.40 \pm 0.52$ \\
M*$-\sigma$\scriptsize(10) & $1.70^{+2.0}_{-2.3}$ & $-0.901^{+1.1}_{-0.94}$ & $2.58^{+1.5}_{-0.74}$ & $-1.06 \pm 0.16$ & $-0.232^{+0.12}_{-0.22}$ & $13.6^{+3.5}_{-3.3}$ & $1.89^{+0.47}_{-0.44}$ & $13.9 \pm 0.22$ & $-1.28^{+0.43}_{-0.4}$ & $5.48^{+2.2}_{-1.6}$ \\
M*$-\sigma$\scriptsize(8) & \cellcolor{red!20}$0$ & \cellcolor{red!20}$0$ & $2.75^{+1.4}_{-0.87}$ & $-1.04 \pm 0.15$ & $-0.263^{+0.14}_{-0.29}$ & $14.4^{+3.2}_{-3.4}$ & $1.85^{+0.48}_{-0.42}$ & $14.0 \pm 0.18$ & $-1.15^{+0.4}_{-0.44}$ & $4.73^{+1.6}_{-1.1}$ \\
M*$-\sigma$\scriptsize(6) & \cellcolor{red!20}$0$ & \cellcolor{red!20}$0$ & \cellcolor{red!20}$2.60$ & \cellcolor{red!20}$-1.00$ & $-0.296^{+0.15}_{-0.25}$ & $15.0^{+2.9}_{-3.7}$ & $1.90^{+0.46}_{-0.39}$ & $14.0 \pm 0.16$ & $-1.09^{+0.35}_{-0.42}$ & $4.94^{+0.81}_{-0.92}$ \\
M*$-\sigma$\scriptsize(4) & \cellcolor{red!20}$0$ & \cellcolor{red!20}$0$ & \cellcolor{red!20}$2.60$ & \cellcolor{red!20}$-1.00$ & \cellcolor{red!20}$-0.28$ & \cellcolor{red!20}$14.0$ & $1.92^{+0.42}_{-0.36}$ & $14.0 \pm 0.16$ & $-1.20^{+0.32}_{-0.36}$ & $4.85 \pm 0.57$ \\
M*$-\sigma$\scriptsize(3) & \cellcolor{red!20}$0$ & \cellcolor{red!20}$0$ & \cellcolor{red!20}$2.60$ & \cellcolor{red!20}$-1.00$ & \cellcolor{red!20}$-0.28$ & \cellcolor{red!20}$14.0$ & \cellcolor{red!20}$1.60$ & $14.0 \pm 0.21$ & $-1.17^{+0.33}_{-0.37}$ & $4.69 \pm 0.55$ \\
M*$-\sigma$\_fgas$-4\sigma$\scriptsize(10) & $1.60^{+2.0}_{-2.2}$ & $-0.863^{+1.0}_{-0.9}$ & $2.61^{+1.3}_{-0.73}$ & $-1.02 \pm 0.13$ & $-0.268^{+0.15}_{-0.34}$ & $13.7 \pm 3.3$ & $1.92^{+0.41}_{-0.36}$ & $14.0 \pm 0.19$ & $-0.906^{+0.29}_{-0.31}$ & $5.64^{+2.2}_{-1.5}$ \\
M*$-\sigma$\_fgas$-4\sigma$\scriptsize(8) & \cellcolor{red!20}$0$ & \cellcolor{red!20}$0$ & $2.89^{+1.3}_{-0.86}$ & $-1.01 \pm 0.12$ & $-0.329^{+0.19}_{-0.41}$ & $14.4^{+3.0}_{-3.3}$ & $1.91^{+0.39}_{-0.35}$ & $14.0 \pm 0.14$ & $-0.814 \pm 0.27$ & $4.81^{+1.4}_{-0.97}$ \\
M*$-\sigma$\_fgas$-4\sigma$\scriptsize(6) & \cellcolor{red!20}$0$ & \cellcolor{red!20}$0$ & \cellcolor{red!20}$2.60$ & \cellcolor{red!20}$-1.00$ & $-0.33^{+0.17}_{-0.27}$ & $15.1^{+2.6}_{-3.4}$ & $1.88^{+0.38}_{-0.34}$ & $14.0 \pm 0.13$ & $-0.809 \pm 0.23$ & $5.28^{+0.72}_{-0.81}$ \\
M*$-\sigma$\_fgas$-4\sigma$\scriptsize(4) & \cellcolor{red!20}$0$ & \cellcolor{red!20}$0$ & \cellcolor{red!20}$2.60$ & \cellcolor{red!20}$-1.00$ & \cellcolor{red!20}$-0.28$ & \cellcolor{red!20}$14.0$ & $1.91^{+0.36}_{-0.3}$ & $14.0 \pm 0.12$ & $-0.867 \pm 0.2$ & $5.16 \pm 0.51$ \\
M*$-\sigma$\_fgas$-4\sigma$\scriptsize(3) & \cellcolor{red!20}$0$ & \cellcolor{red!20}$0$ & \cellcolor{red!20}$2.60$ & \cellcolor{red!20}$-1.00$ & \cellcolor{red!20}$-0.28$ & \cellcolor{red!20}$14.0$ & \cellcolor{red!20}$1.60$ & $14.0 \pm 0.15$ & $-0.852 \pm 0.2$ & $5.05 \pm 0.5$ \\
\end{longtable}

\normalsize

\end{landscape}
\restoregeometry
\clearpage